\documentclass[aps,prl,twocolumn,showpacs,superscriptaddress]{revtex4-1}

\usepackage{amssymb,amsfonts,amsmath}
\usepackage{nicefrac}

\usepackage[pdftex]{graphicx}
\usepackage{graphicx}
\usepackage{color}
\usepackage{dcolumn}
\usepackage{multirow}
\usepackage{hyperref}

\def\simlt{\mathrel{\lower .3ex \rlap{$\sim$}\raise .5ex \hbox{$<$}}}
\def\simgt{\mathrel{\lower .3ex \rlap{$\sim$}\raise .5ex \hbox{$>$}}}

\newcommand{\ket}[1]{ |{#1} \rangle }

\newcommand{\snc}[1]{\textcolor{black}{#1}}
\newcommand{\suec}[1]{\textcolor{black}{#1}}

\begin{document}

\title{High fidelity gates in quantum dot spin qubits}

\author{Teck Seng Koh}
\author{S. N. Coppersmith}
\author{Mark Friesen}
\affiliation{Department of Physics, University of Wisconsin-Madison, Madison, WI 53706, USA}


\begin{abstract}

Several logical qubits and quantum gates have been proposed for semiconductor quantum dots controlled by voltages applied to top gates.
\suec{The different schemes can be difficult to compare meaningfully.}
Here we develop a theoretical framework to evaluate disparate qubit-gating schemes on an equal footing.
We apply the procedure to two types of double-dot qubits:  the singlet-triplet (ST) and the semiconducting quantum dot hybrid qubit.  
We investigate three quantum gates that flip the qubit state:  a DC pulsed gate, an AC gate based on logical qubit resonance (LQR), and a gate-like process known as stimulated Raman adiabatic passage (STIRAP).
These gates are all mediated by an exchange interaction that 
\suec{is controlled experimentally using}
the interdot tunnel coupling $g$ and the detuning $\epsilon$, which sets the energy difference between the dots.
Our procedure has two steps.
First, we optimize the gate fidelity ($f$) for fixed $g$ as a function of the other control parameters;
this yields an $f^\text{opt}(g)$ that is universal for different types of gates.
Next, we identify physical constraints on the control parameters; this yields an upper bound $f^\text{max}$ that is specific to the qubit-gate combination.
We show that similar gate fidelities ($ \sim 99.5$\%) should be attainable for ST qubits in isotopically purified Si, and for hybrid qubits in natural Si.
Considerably lower fidelities are obtained for GaAs devices, due to the fluctuating magnetic fields $\Delta B$ produced by nuclear spins.

\end{abstract}

\maketitle



\begin{figure*}[t]
\begin{center}
\vspace{-.2cm}
\centerline{\includegraphics[scale=0.28]{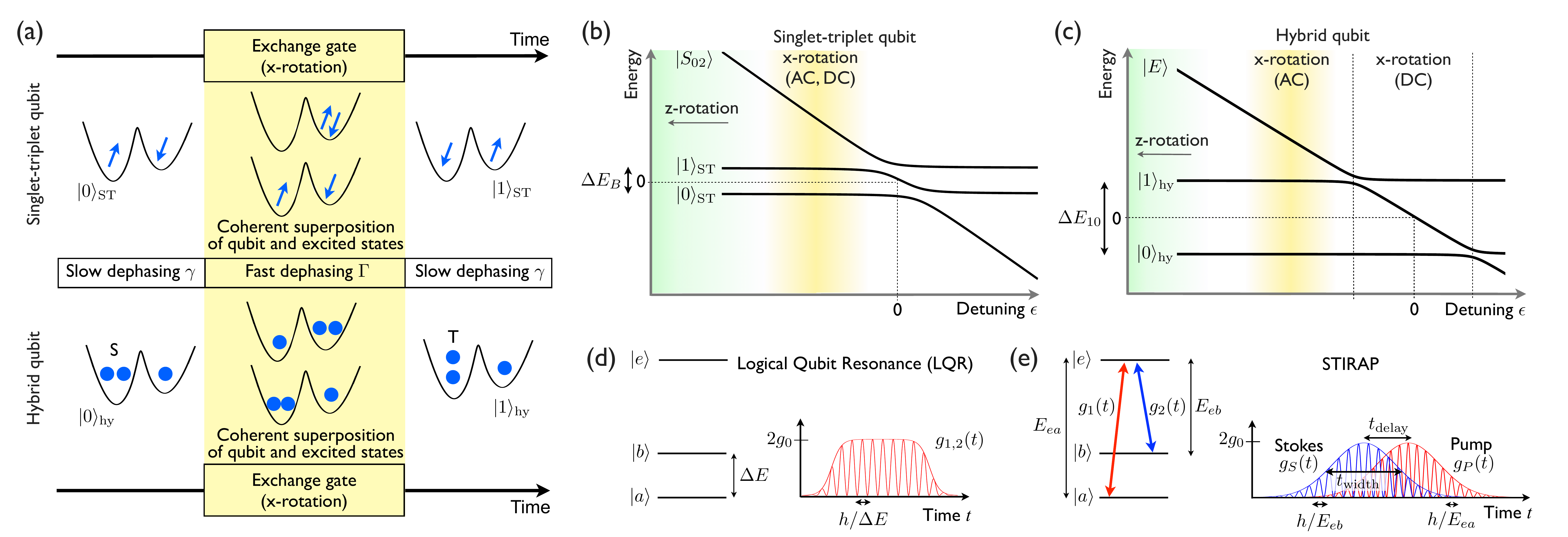}}
\caption{\label{fig:cartoon} 
Physics of $x$ and $z$-rotations of singlet-triplet (ST) and hybrid qubits in a double quantum dot.
Transitions between qubit states yield $x$-rotations, while 
\suec{$z$-rotations change the relative phase of the qubit states.}
(a) Schematics of processes underlying $x$-rotations for  ST and hybrid qubits, which are implemented by changing the detuning
to turn on an exchange interaction by mixing in an excited state with different charge character. 
The qubit states are shown on the left and right-hand sides,
while the intermediate states are shown in the middle of the panel.
Increasing the strength of the exchange interaction increases the gate speed but also increases  dephasing  from charge fluctuations.
Reducing the gate speed increases the exposure to spin dephasing.
In this paper, we determine the optimal gate speeds that maximize the gate fidelity.
(b), (c) The three lowest energy levels for ST and hybrid qubits, including the two qubit states and the excited charge state, as a function of the detuning, $\epsilon$.
The energy splitting of the qubit states drives $z$-rotations, and is typically much smaller for ST qubits than for hybrid qubits (\emph{i.e.}, $\Delta E_B \ll \Delta E_{10}$). 
For a given value of the tunnel coupling $g$, the exchange interaction $J$ is large when $|\epsilon|$ is small.  Large $J$ causes faster gate speed but also faster decoherence from charge noise.
For ST qubits, optimum fidelity is obtained when  $\epsilon <0$.
For hybrid qubits, AC gates are optimized with $\epsilon < 0$, while DC pulsed gates are optimized when $\epsilon$ is set to the energy level anticrossings~\cite{Koh:2012p250503}.
(d) Logical qubit resonance (LQR) is performed by oscillating the tunnel coupling $g(t)$ [and therefore $J(t)$] at the primary resonant frequency $\omega=\Delta E/2\hbar$, or at the secondary harmonic $\omega=\Delta E/\hbar$.  
(The secondary period is identified in the figure.)
(e) Stimulated Raman rapid adiabatic passage (STIRAP)~\cite{Shore:1990} is implemented at the resonant frequencies, $\omega =E_{ea}/\hbar$ and $E_{eb}/\hbar$, using overlapping Gaussian envelopes known as Stokes and pump pulses.}
\end{center}
\vspace{-.1cm}
\end{figure*}

{T}he fundamental building block of a quantum information processor is a two-state quantum system, or qubit.  
Solid state qubits based on electrons confined in top-gated quantum dots in semiconductor heterostructures~\cite{Loss:1998p120} are promising, due to the promise of manipulability, and the overall maturity of semiconductor technology.
\snc{In a charge qubit, the information is stored in the location of an electron in a double quantum dot.}
Because charge qubits are subject to strong Coulomb interactions, they can be manipulated quickly,
at gigahertz frequencies, using control electronics~\cite{Hayashi:2003p226804, Gorman:2005p090502, Petersson:2010p246804,Shi:2012preprint}.
{\color{black}
However, they also couple strongly to environmental noise sources, such as thermally activated charges on materials defects, }
leading to short, sub-nanosecond decoherence times~\cite{Dovzhenko:2011p161302}.
\suec{Spin qubits, which couple more weakly to environmental noise, have much longer coherence
times}~\cite{Levy:2002p147902, Petta:2005p2180, Shi:2012p140503,Loss:1998p120, Koppens:2006p766, DiVincenzo:2000p339, Laird:2010p075403, Maune:2012p344,Pla:2012p541,Tyryshkin:2012p143}.
However, 
because magnetic couplings are weak, gate operations between spin qubits are slow.
For this reason, in most gating protocols spin qubits adopt a charge character briefly during gate operations.
Successful gate operations generally entail a 
tradeoff: charge-like for faster gates vs.\ spin-like for better coherence.

Several types of logical qubits have been designed to enable electrically controlled
manipulation and measurement of qubits encoded in spin degrees
of freedom formed of 
{\color{black}
two or more electrons in two~\cite{Levy:2002p147902, Taylor:2005p177,Shi:2012p140503} or three~\cite{DiVincenzo:2000p339} coupled dots.}
These logical qubits share 
experimental control knobs; however, their spin-charge characteristics vary widely, yielding variations in gating speeds, dephasing rates, and 
gating protocols.

When characterizing quantum gates, instead of considering
the gating time and decoherence time separately, 
it is important
to consider the gate fidelity, 
a measure of the fraction of the wavefunction that is in the targeted state, which
depends on the ratio of the gating time to the decoherence time.
Here we argue that achieving meaningful comparisons between logical qubits and gating schemes is greatly facilitated by first optimizing the specific gate operations, taking into account the different dephasing rates of the spin and charge sectors.
We compute and optimize gate fidelities for different qubits and gating protocols
using a master equation approach.

We consider two types of logical qubits in a double quantum dot:  a singlet-triplet (ST) qubit formed with two electrons, one in each dot~\cite{Levy:2002p147902, Petta:2005p2180}, and a quantum dot
hybrid qubit formed with three electrons, two in one dot and one in the other~\cite{Shi:2012p140503,Ferraro:2013p1304.1800}.
{\color{black}
Logical qubit states for the ST qubit are
$\ket{0}_\text{ST}$=$\ket{\uparrow}_L\ket{\downarrow}_R$ and
$\ket{1}_\text{ST}$=$\ket{\downarrow}_L\ket{\uparrow}_R$, where $\ket{\uparrow}$ and $\ket{\downarrow}$ are spin up or down states, and $L$ and $R$ refer to the left or right dots.
Logical qubit states of the quantum dot hybrid qubit 
are $\ket{0}_\text{hy}$=$\ket{S}_L\ket{\downarrow}_R$ and
$\ket{1}_\text{hy}$=$\sqrt{\nicefrac{1}{3}}\ket{T_0}_L \ket{\downarrow}_R
- \sqrt{\nicefrac{2}{3}}\ket{T_{-}}_L\ket{\uparrow}_R$, where
$\ket{S},\ket{T_0}$=$\sqrt{\nicefrac{1}{2}}(\ket{\uparrow\downarrow}\mp \ket{\downarrow\uparrow})$ and
$\ket{T_-}$=$\ket{\downarrow\downarrow}$
are singlet (S) and triplet (T) states.}
{\color{black} 
Energy differences between the qubit states drive $z$-rotations around the Bloch sphere.  
For ST qubits, the energy splitting $\Delta E_B$ is caused by a magnetic field difference $\Delta B$ on the two sides of the double dot.  
$\Delta B$ occurs naturally in GaAs and natural Si, and may be enhanced by nuclear polarization~\cite{Petta:2008p067601, Foletti:2009p903}, or 
with
micromagnets~\cite{PioroLadriere:2008p776} or striplines~\cite{Koppens:2006p766}.
Typical values of $\Delta B$ are in the range $10^{-6}$-$10^{-2}$~T.
Hybrid qubits do not require local magnetic fields; the qubit energy splitting $\Delta E_{10}$ is dominated by the singlet-triplet energy splitting of the two-electron dot.
$\Delta E_{10}$ is typically of order $0.1 \text{ meV}$~\cite{Shi:2011p233108}, yielding much faster
$z$-rotations than in ST qubits.
 
Although $z$-rotations can never be extinguished in ST or hybrid qubits, the rotation axis may be varied in the $x$-$z$ plane by adjusting the tunnel coupling between the two sides of the double dot.}
As Fig.\ \ref{fig:cartoon} indicates, the main experimental parameters are $g$ and the detuning $\epsilon$, which characterizes the energy difference between different charge configurations [(1,1) vs.\ (0,2) for the ST and (2,1) vs.\ (1,2) for the hybrid qubit].
We use analytical and numerical calculations to find the relationship between $\epsilon$ and $g$ that maximizes the fidelity $f(\epsilon,g)$ of $x$ and $z$-rotations.
Physical limits on $\epsilon$ and $g$ for a given qubit scheme then determine the maximum achievable fidelity.

We consider three different schemes for 
performing $x$-rotations: 
(i) DC pulsed gates~\cite{Petta:2005p2180}, in which the detuning is changed suddenly between different values; 
(ii) logical qubit resonance (LQR), an AC resonant technique analogous to electron spin resonance (ESR) for single spins~\cite{Poole:1983}, and 
(iii) stimulated Raman adiabatic passage (STIRAP)~\cite{Bergmann:1998p1003}, another AC resonant technique in which each qubit state is coupled to an auxiliary excited state.
Given a tunnel coupling $g$, pulse-gating and LQR are optimized over the detuning $\epsilon$,
while STIRAP is optimized over the duration of the pulses used.
Remarkably, we find that the $g$-dependence of the optimal fidelity $f^\text{opt}(g)$ is very similar 
for all three gating schemes.
However, physical constraints that differ between the gating schemes limit the achievable fidelity $f^\text{max}$.

The paper is organized as follows.  
The next section provides relevant details concerning ST and hybrid qubits and their decoherence rates.
We describe the physical mechanisms for implementing $x$-rotations (transitions between qubit states) and $z$-rotations (changes in the phase difference between the qubit states).
We discuss the 
``slow" or ``pure" spin dephasing rate $\gamma$, arising from dephasing of the qubit states themselves, and the ``fast" charge dephasing rate $\Gamma$, involving the intermediate state~\cite{Hu:2006p100501,Gamble:2012p035302}.
We then present the Calculations and Results for qubit fidelities, based on the master equations presented in Methods. (Additional details are provided in the Appendix.)
Figs.\ \ref{fig:acfidelities} and \ref{fig:DCfidelities} show the key results,
plots of optimized fidelities as a function of 
the tunnel coupling $g$.
The Discussion describes the physical constraints that determine the upper bounds on $f^\text{opt}$, for each type of qubit, gate operation, and materials system (Si vs.\ GaAs).

\section{Logical Qubits, Gates, and Decoherence Mechanisms}
Fig.~\ref{fig:cartoon} shows gating schemes and energy levels for 
ST and hybrid 
quantum dot qubits.
The horizontal energy levels in the $\epsilon<0$ portion of Figs.\ \ref{fig:cartoon}(b) and (c) correspond to the logical qubit states.
Only states that can be reached by \suec{spin-conserving} processes 
are shown.
A third state with a different charge configuration that plays a prominent role during gating
is shown for both ST and hybrid qubits ($|e\rangle=|S_{02}\rangle$ or $|E\rangle$, respectively).
At (or near)
the detuning value $\epsilon=0$, states with different charge configurations 
are energetically degenerate [(1,1) and (0,2) states for the ST qubit,
 (2,1) and
(1,2) states for the hybrid qubit].
\suec{We} focus on the regime $\epsilon\leq 0$.

{\em Implementations of $x$-rotations.}
The implementations of $x$-rotations for
ST and hybrid qubits discussed here involve the exchange interaction, which is mediated by
the excited state $|e\rangle$.
Figure~\ref{fig:cartoon}(a) demonstrates the exchange process for ST and hybrid qubits.
Decreasing $|\epsilon|$ increases the occupancy of $\ket{e}$, which
enhances the speed of $x$-rotations, but also increases the coupling to external charge noise~\cite{Barrett:2002p125318}.
For both ST and hybrid qubits, the rate $\Gamma$ of charge dephasing between $\ket{e}$
and the qubit states
is much faster than the rate $\gamma$ of pure dephasing  between the qubit states, 
so changing $\epsilon$ strongly affects the gate fidelity.
Charge noise couples to both $\epsilon$ and $g$, yielding distinct dephasing mechanisms~\cite{Taylor:2007p035315}.
However, as shown below, the highest fidelities are obtained when $\epsilon \ll 0$, in the region where fluctuations in $g$ are dominant, because the qubit energy levels have 
very nearly the same dependence
on detuning~\cite{footnote1}
Therefore, we consider only $g$-noise here.

\emph{DC Pulsed Gates.}
ST qubit experiments typically keep the tunnel coupling fixed and use $\epsilon$ to tune the exchange coupling~\cite{Petta:2005p2180, Maune:2012p344}, 
as indicated in Fig.\ \ref{fig:cartoon}.
$z$-rotations are obtained when $J(\epsilon)\ll \Delta E$, while $x$-rotations are obtained when $J(\epsilon)\gg \Delta E$.
In pulsed gating protocols, $\epsilon$ is switched between these two positions 
quickly, so that the quantum state does not evolve significantly during the switching time.

In a hybrid qubit, the energy splitting between the qubit states is much larger than the tunnel couplings ($\Delta E_{10} \gg g$). 
The energy level diagram then has two distinct anticrossings, as indicated by vertical dotted lines in Fig.\ \ref{fig:cartoon}(c). 
The pulse-gating scheme proposed in~\cite{Koh:2012p250503} to implement an arbitrary rotation on the Bloch sphere 
has five steps, three of which are at anticrossings.
Below, we show that this requirement leads to serious constraints on gate fidelities 
using current technology.

\emph{Logical Qubit Resonance (LQR).}
In conventional ESR~\cite{atherton1993principles}, a 
DC magnetic field applied along $\hat{\mathbf{z}}$
induces a Zeeman splitting, $\Delta E$.  
A small AC transverse magnetic field applied 
along $\hat{\mathbf{x}}$
at the resonant frequency $\omega=\Delta E/\hbar$ induces transitions between states with different values of spin component $S_z$.
In the analogous LQR scheme [Fig.\ \ref{fig:cartoon}(d)], the qubit energy splitting ($\Delta E=\Delta E_B$ or $\Delta E_{10}$) plays the role of the Zeeman energy, 
while an oscillating exchange interaction $J(t)$ plays the role of the transverse field.
As for pulsed gates, $\epsilon$ is increased from a value $\ll$ 0 to a value closer to zero, 
where an increase in tunnel coupling $g$ creates a significant exchange interaction.
$\epsilon$ is then held constant, while
$g(t)=g_0[1+\cos (\omega t)]$ 
is modulated.
Here, we assume that $g$ oscillates between zero and a positive value $2g_0$,
as indicated in Fig.\ \ref{fig:cartoon}(d).
The amplitude of the AC component of 
$J(t)$
determines the speed of the $x$-rotation.

LQR differs from conventional ESR in two main ways.
First, since the oscillating exchange interaction $J$ has a nonzero DC component ($J_\text{DC}$), the precession axis tilts slightly away from $\hat{\bf z}$.
Within the rotating wave approximation discussed in the Appendix, 
this leads to infidelity in the LQR gate because
the effective $B$-field has a shifted magnitude ($\sqrt{\Delta E^2+J_\text{DC}^2}$).
(This error is accounted for in all the calculations shown here.)
Second, the primary resonance occurs at \emph{half} the Larmor frequency $\omega=\Delta E/2\hbar$.
{\color{black}
This is because the tunnel coupling, $g(t)\sim 1+\cos (\omega t)$, generates two different AC components.
For example, $J\sim g^2/U$ yields a primary component, $\sim\cos (2\omega t)/2$, and a secondary component, $\sim 2\cos (\omega t)$.  (See Appendix.)}
The numerical results reported here all correspond to the secondary resonance, since it yields slightly higher fidelities.

\begin{figure}[t]
\begin{center} 
\vspace{-.1in}
\centerline{\includegraphics[scale=0.32]{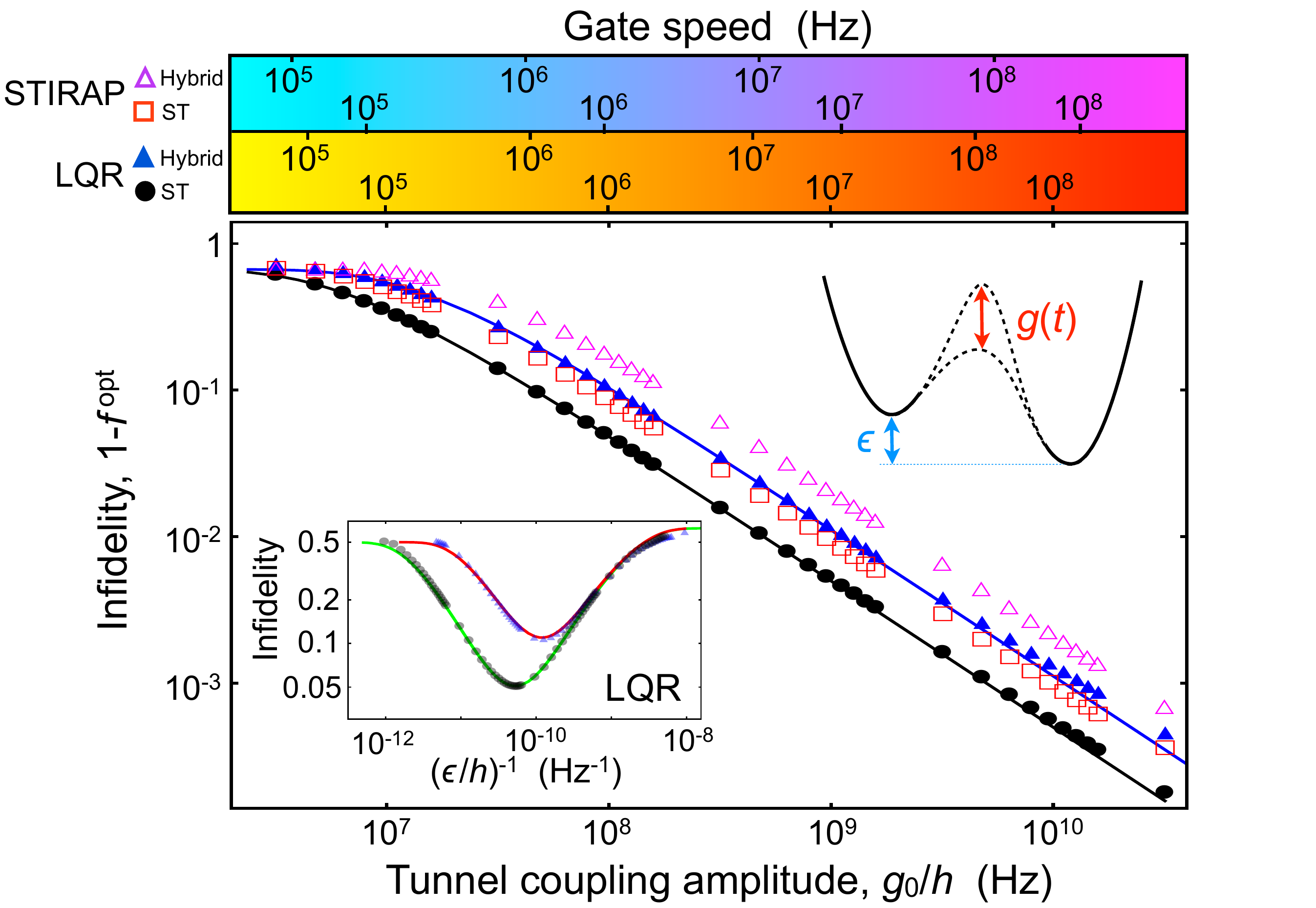}}
\caption{\label{fig:acfidelities} 
Fidelities for LQR and STIRAP~\cite{footnote2} $x$-rotations in ST and hybrid qubits, for isotopically pure $^{28}$Si
 (see legend at top left), \suec{using material parameters given in the main text.}
We assume a magnetic field {\color{black}difference} of $\Delta B=0.3$~mT, from an external micromagnet.
Lower inset:  
\suec{at fixed $g_0$, 
the optimized value 
$\epsilon^\text{opt}(g_0)$ is found by minimizing the infidelity $1$-$f$ as a function of $\epsilon$.}
Data points are numerical results for $g_0/h=0.3/\pi$~GHz\suec{,}
\suec{and} solid lines are analytical estimates.  (See Appendix.)
\suec{The numerically obtained
$\epsilon_\text{LQR}^\text{opt}$ 
agree with the analytical 
estimate,}
Eq.\ \eqref{eq:optimaldetst}. 
Main panel:  \suec{data points are numerically optimized fidelities $f^\text{opt}=f(\epsilon^\text{opt})$,
versus} $g_0$. 
Solid lines are analytical estimates for $f_\text{LQR}^\text{opt}$, Eq.\ \eqref{eq:optfidelityhy}.
The corresponding gate speeds increase with $g_0$, as indicated in the top panels. 
{\color{black}
Upper inset:  a cartoon showing the tunnel barrier and detuning in a double dot,
which are both controlled electrostatically.}}
\end{center}
\end{figure}

\emph{Stimulated Raman Adiabatic Passage (STIRAP).}
The STIRAP protocol~\cite{Bergmann:1998p1003} generates $x$-rotations on the Bloch sphere by inducing transitions between the qubit states $|a\rangle$ and $|b\rangle$.
A simple STIRAP protocol is shown in Fig.\ \ref{fig:cartoon}(e).
The tunneling processes $|a\rangle\leftrightarrow|e\rangle$ and $|b\rangle\leftrightarrow|e\rangle$ are controlled independently by oscillating $g(t)$ at the resonant frequencies $\hbar \omega_P=\Delta E_{ea}$ and $\hbar \omega_S = \Delta E_{eb}$.
Again, we assume the tunnel coupling is non-negative, with a DC component $g_0$, and an AC amplitude $2 g_0$.
Counterintuitively, an adiabatic pulse sequence with $g_2(t)=g_S(t)$ followed by $g_1(t)=g_P(t)$ produces a rotation from $|a\rangle$ to $|b\rangle$ that never populates $|e\rangle$, and therefore never experiences charge dephasing.
Realistic pulse sequences have a finite duration however; this yields a small population of $|e\rangle$, and therefore dephasing. 
Similar to pulsed gates and LQR, we anticipate there will be an optimal gate speed that maximizes the process fidelity.
{\color{black}
We note that the standard STIRAP protocol shown in Fig.\ \ref{fig:cartoon}(e) is not a true qubit gate~\cite{footnote2}.
True gates can be achieved by employing longer, STIRAP-like pulses~\cite{Lacour:2006p362},
which must be optimized over many more parameters.  We only study the standard, two-pulse sequence here, to focus on the fundamental physics limiting the fidelity of the protocol.
}

{\em Decoherence mechanisms.}
The key physics incorporated in our calculations is that the qubit states, which have a spin
character, often have a much lower dephasing rate than the states accessed during a gate
operation, which typically have a substantial charge character~\cite{Hu:2006p100501}.
To understand the achievable fidelities of real devices, our calculations use experimentally
realistic numbers, which we list here.

Charge qubit experiments indicate that the fast charge noise dephasing rate $\Gamma$ is very similar in Si and in GaAs~\cite{Petersson:2010p246804, Hayashi:2003p226804, Shi:2012preprint}.
Here, we adopt the value $\Gamma = 1$~GHz.
The much slower pure dephasing rate $\gamma$
depends significantly on the material host and the type of qubit.
For ST qubits, pure dephasing is caused by the slow diffusion of nuclear spins. 
We adopt the values $\gamma_\text{ST}=0.2$~MHz for 99.99\% isotopically purified $^{28}$Si, 4.5~MHz for natural Si, and 0.14~GHz for GaAs, which are obtained as quadrature sums of contributions from the nuclear hyperfine coupling~\cite{Assali:2011p165301} and the electron-phonon coupling~\cite{Gamble:2012p035302,Hu:2011p165322}.
For hybrid qubits, we use $\gamma_\text{hy}=1$~MHz for 99.99\% isotopically purified $^{28}$Si, 4.6~MHz for natural Si, and 5.9~GHz for GaAs, with the main contributions to dephasing coming from charge noise and optical phonons~\cite{Gamble:2012p035302,Hu:2006p100501}.

While application of
echo sequences can be used to greatly increase the coherence times of quiescent qubits~\cite{Bluhm:2011p109} and of $z$-rotations,
{\color{black}
it is nontrivial to correct for low-frequency noise during a gate sequence.
While several correction schemes have been proposed~\cite{Khodjasteh:2009p080501,Wang:2012p997} and noise suppression schemes have been implemented~\cite{Bluhm:2010p216803},
the required pulse sequences are rather complicated.
Here, we  only study short sequences, so the dephasing rates in our calculations must include the low-frequency noise.}

\section{Calculations and Results}
We now present our results for the optimized fidelity of single-qubit gate operations in the presence of both fast and slow dephasing mechanisms.
We first focus on the fidelity of $x$-rotations.
As described in Methods, we solve a master equation for the density matrix $\rho$.
For both ST and hybrid qubits,
the coherent evolution is governed by a three-state Hamiltonian, $H$, involving the two logical qubit states, $|a\rangle$ and $|b\rangle$, and the excited charge state, $|e\rangle$, with
 a fast dephasing rate $\Gamma$
between the excited state and each of the qubit states, and
a slow dephasing rate $\gamma$ between the qubit states.
The pulsed, LQR, and STIRAP protocols are implemented by modulating the detuning
$\epsilon (t)$ and the tunnel coupling $g(t)$. 
Dephasing is introduced through a Markovian phenomenological term $D$~\cite{Gardiner:2004} that incorporates dephasing associated with charging transitions in a double quantum dot~\cite{Barrett:2002p125318}.

We present the fidelities of different gating schemes for the specific gate operation of 
a $\pi$-rotation about the $x$-axis 
from the initial state $|a\rangle$ [initial density matrix $\rho_{aa}(0)=1$]
to the final state $|b\rangle$ [target density matrix $\rho_{bb}(\tau)=1$], for a gate
that is implemented in a time $\tau$.
Our fidelity measure is the distance between the actual and ideal density matrices for a $\pi$-rotation~\cite{Jozsa:1994p2315}, which is the calculated value of $\rho_{bb}(\tau)$.  (See Appendix.)
For pulsed gates, we consider a one-step pulse sequence for ST qubits~\cite{Petta:2005p2180, Maune:2012p344}, and a five-step sequence for hybrid qubits~\cite{Koh:2012p250503}.
For the AC gates, we solve the master equation 
within the rotating wave approximation (RWA)~\cite{Shore:1990}.

We first optimize the LQR gates.
For a fixed value of $g_0$, the value of $\epsilon$ at which the fidelity $f$ is maximized, $\epsilon^\text{opt}(g_0)$, is found.
(The inset of Fig.\ \ref{fig:acfidelities} shows the infidelity 1-$f$, which exhibits a minimum.) 
For small  $|\epsilon|^{-1}$ (large detuning), the gate speed is slow and the fidelity is limited by the pure dephasing rate $\gamma$. 
For large $|\epsilon|^{-1}$, the gate speed is fast and the fidelity is limited by the charge noise dephasing rate $\Gamma$.
The optimum fidelity, which is achieved at the crossover between the two regimes, is 
determined numerically.
The Appendix presents the derivation of
analytical estimates for the fidelity as a function of $\epsilon$ (solid lines in the lower inset of Fig.\ \ref{fig:acfidelities}) and of the optimal detuning and fidelity for LQR gates driven at the secondary resonance 
(solid lines in the main panel of Fig.\ \ref{fig:acfidelities}):
\begin{equation}
\left | \epsilon_\text{LQR}^\text{opt} 
\right | \simeq g_0 \sqrt{8\Gamma/\gamma} , \label{eq:optimaldetst}
\end{equation} 
and
\begin{equation}
f^{\text{opt}}_\text{LQR} \simeq 1-(\pi\hbar/g_0)\sqrt{\Gamma \gamma/2} 
\simeq \frac{1}{3}+\frac{2}{3}e^{ -(3h/4g_0)\sqrt{\Gamma \gamma/2} }
\label{eq:optfidelityhy} .
\end{equation}  
Numerical results are also shown.
Results for the fidelity of LQR at the primary resonance 
can be obtained by replacing $g_0\rightarrow g_0/2$ in Eqs.~\eqref{eq:optimaldetst} and \eqref{eq:optfidelityhy}, yielding a lower fidelity.

\begin{figure}[t]
\begin{center}
\centerline{\includegraphics[scale=0.22]{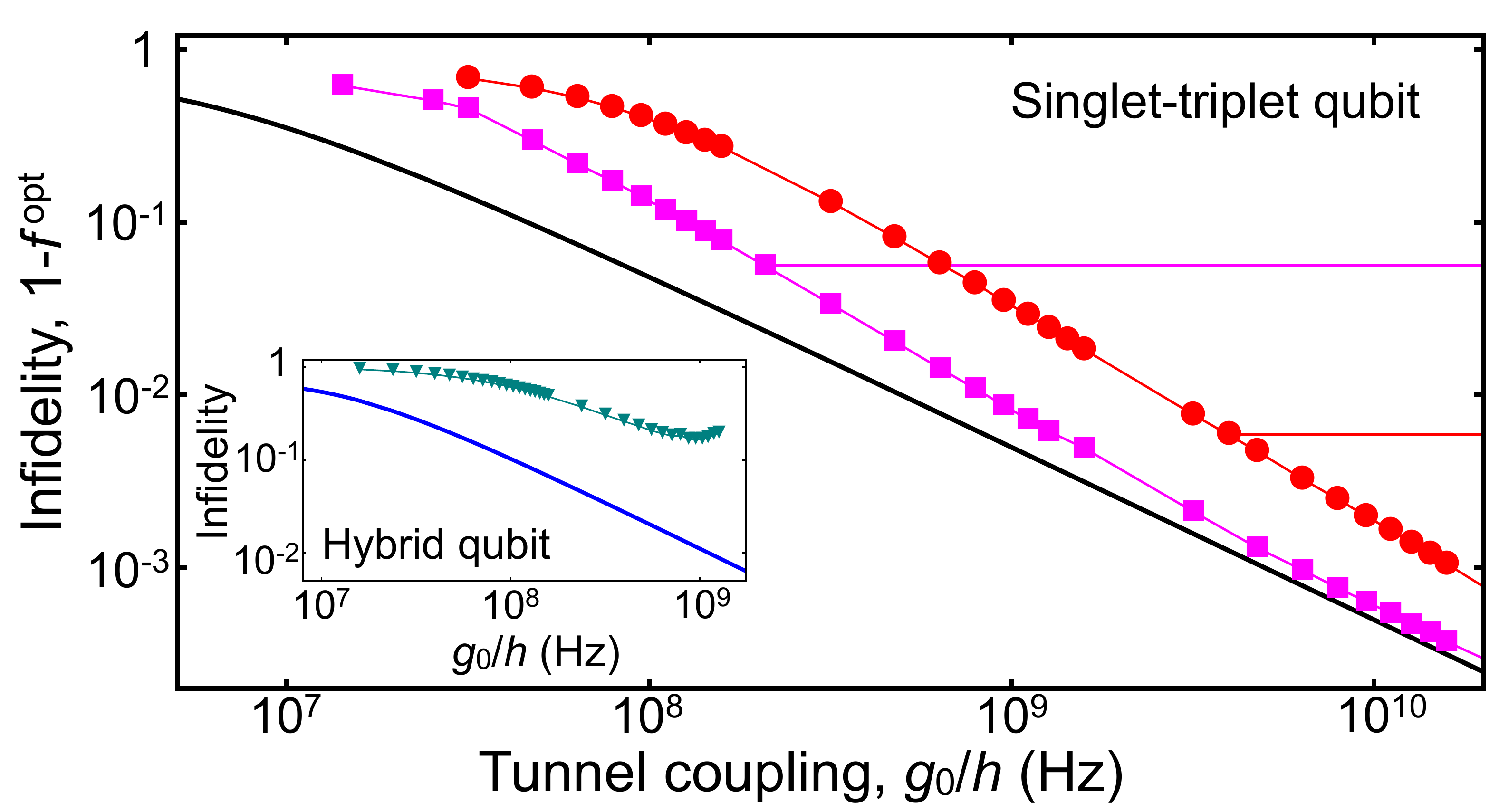}}
\caption{\label{fig:DCfidelities}
Optimized fidelities for pulsed gate $x$-rotations in ST and hybrid qubits, for isotopically \suec{pure} $^{28}$Si.
Main panel:  numerical results for the infidelity vs.\ tunnel coupling $g_0$ for an ST qubit with 
$\Delta B = 0.03$~mT (magenta squares) and $\Delta B = 0.3$~mT (red circles). 
The solid black line \suec{is} an upper bound on the fidelity, obtained when $\Delta E_B=0$.
The analytical form is the same as for LQR gates in ST qubits at the secondary resonance (solid black curve in Fig.\ \ref{fig:acfidelities}).
The numerical results deviate from this limiting behavior most significantly at small $g_0$, in the regime where $J\lesssim \Delta E_B$ and the rotation axis points away from $\hat{\bf x}$. 
The horizontal lines describe the infidelity of $z$-rotations for the same \suec{$\Delta B$}.
Inset:  numerical results for the infidelity of a pulse-gated hybrid qubit, using the five-step pulse sequence described in~\cite{Koh:2012p250503} (green triangles).
As a comparison, the solid line is the analytical estimate for an LQR gate in a hybrid qubit at the secondary resonance (solid blue curve in Fig.\ \ref{fig:acfidelities}).
Pulsed gates in hybrid qubits have relatively poor fidelity because $\epsilon$ cannot be optimized.
}
\end{center}
\end{figure}

Pulsed gates in ST qubits are optimized similarly to LQR gates, yielding similar results.
Fig.\ \ref{fig:DCfidelities} shows numerically optimized fidelities
for two different interdot magnetic field differences, $\Delta B$.
In the low-field regime $\Delta E_B\ll J$, the rotation axis points nearly along $\hat{\bf x}$.
When $\Delta B=0$, we can obtain analytical estimates for the optimized detuning and fidelity, obtaining the same results as Eqs.~\eqref{eq:optimaldetst} and \eqref{eq:optfidelityhy}, with
$\epsilon_\text{LQR}^\text{opt} \rightarrow  \epsilon^\text{opt}_\text{ST,DC}$ and
$f^\text{opt}_\text{LQR} \rightarrow f^\text{opt}_\text{ST,DC} $.  (See Appendix.) 
Fig.\ \ref{fig:DCfidelities}, we see that the numerically optimized fidelities approach this limiting behavior for large $g_0$ or small $\Delta B$.
For smaller $g_0$, the fidelity is suppressed by a combination of dephasing effects, and a misalignment of the rotation axis from $\hat{\bf x}$. 
A three-step pulse sequence that corrects the rotation angle~\cite{Hanson:2007p050502}
yields only small improvements in the fidelity.  (See Appendix.)

Pulsed gates in hybrid qubits differ significantly from the other gating schemes because the optimal value of $\epsilon$ does not depend on $g_0$.
To understand this, we note that a general, pulsed gate rotation sequence for a hybrid qubit requires five steps~\cite{Koh:2012p250503}, with three
 of these steps occurring at anticrossings.
Dephasing errors are minimized by maximizing the transition speed, \emph{i.e.}, by tuning $\epsilon$ directly to the level anticrossings in Fig.\ \ref{fig:cartoon}(c).
This yields the results shown in the inset of Fig.\ \ref{fig:DCfidelities}.
The inset also shows
 the optimal fidelity for an LQR gate in a hybrid qubit;
LQR typically achieves a much higher fidelity.

We next present results for optimized fidelities of the STIRAP~\cite{footnote2} protocol for ST and hybrid qubits.
The pulse shape determines
the gate speed of STIRAP [Fig.\ \ref{fig:cartoon}(e)].
For a given value of $2g_0$,
the pulse shape parameters $t_\text{width}$ and $t_\text{delay}$ are optimized for maximum fidelity.
There are no simple analytical methods for treating the STIRAP protocol, so the optimal fidelities are obtained numerically, yielding the results shown in Fig.\ \ref{fig:acfidelities}.
Remarkably, we find that the optimal fidelities for STIRAP and LQR gates exhibit the same dependence on $g_0$, differing only by a small factor.  (See Appendix.)
The various gate speeds are indicated by the calibration bars at the top of Fig.\ \ref{fig:acfidelities}.

The analysis of the fidelity of $z$-rotations is considerably simpler than the analysis
of $x$-rotations presented above.
For both ST and hybrid qubits,
the fidelity of a $z$-rotation should be understood simply as a competition between the pure dephasing rate $\gamma$, and the gate speed, where the latter is determined by the energy splitting $\Delta E$ between the qubit states.

\section{Discussion}
The previous section presents relations between the control parameters that yield optimized gate fidelities $f^\text{opt}(g_0)$.
{\color{black}
In principle, the fidelity can be made arbitrarily close to $1$ by increasing $g_0$.
In practice, physical constraints on the experimental control parameters
bound the fidelity.
We now list the constraints that bound the fidelity of $x$-rotations.

\emph{Level spacing.}  
The anticrossings in the qubit energy level diagram should be well separated, otherwise transitions may occur between \emph{three} levels rather than two.
$g_0$ should therefore be smaller than the spacing between single-particle levels in the dots.  
Recently, this condition was found to be satisfied in an electrostatically-defined SiGe double dot for which $g_0\simeq 40$~$\mu$eV \cite{Simmons:2009p3234}.
We therefore assume a bound of $g_0/h<10$~GHz for systems of this type.
For hybrid qubits, $g_0$ should also be smaller than the qubit energy splitting, or $g_0<\Delta E_{01}/2$.
When this constraint is not satisfied, the infidelity rises, as on the right-hand-side of the inset of Fig.\ \ref{fig:DCfidelities}.

\emph{RWA.}
Resonant gating requires that many fast, resonant oscillations fit inside a single pulse envelope. 
This is the basis for the RWA, and it yields the following constraints  (see Appendix):  $g_0\ll \sqrt{|\epsilon| \Delta E}/2$ for the secondary resonance of LQR, and $g_0\ll 2|\epsilon|/\sqrt{\pi}$  for STIRAP.
When $\epsilon =\epsilon^\text{opt}(g_0)$, the LQR requirement further simplifies to $g_0\ll\Delta E\sqrt{\Gamma/2\gamma}$.
For ST qubits, the constraint is quite strict and yields relatively low fidelities.
Our numerical calculations suggest that in this situation the fidelity can be improved by deviating from 
Eq.\ \eqref{eq:optimaldetst}.
For the STIRAP scheme, there is no analogous relation between $\epsilon^\text{opt}$ and $g_0$, and the RWA is far less restrictive.

\emph{Adiabaticity.}
Pulsed gating methods require instantaneous pulses.
However, \suec{rise times} in real experiments 
are finite.
\suec{For pulsed gates, the evolution is effectively instantaneous when the time dependence of the energy difference at an anticrossing $\Delta(t)$ satisfies} $g_0^2\ll \hbar \, (d\Delta/dt)$~\cite{Vutha:2010p389}. 
Using experimental measurements and numerical calculations, and assuming a realistic rise time of $\sim$100~ps for currently available pulse generators~\cite{Petersson:2010p246804,Shi:2012preprint}, we deduce a bound of $g_0/h \ll 3$~GHz for pulsed gates.

\emph{Misorientation.}
If $z$-rotations cannot be turned off, the fidelity of an $x$-rotation will be limited by the misorientation of the rotation axis.
This is always true for simple pulsed gates in hybrid and ST qubits because of the energy splitting between the qubit states.
For ST qubits, the problem is mitigated by reducing the magnetic field difference $\Delta B$ (e.g., by modifying the micromagnets), or by increasing $g_0$ (and therefore the $x$ component of the rotation).
In the latter case, the fidelity can be improved by increasing both $\epsilon$ and the charging energy, as described below.
Alternatively, a three-step pulse sequence can be used to correct the misorientation~\cite{Hanson:2007p050502}.
For hybrid qubits, the problem is more severe and a single-step sequence is untenable~\cite{Koh:2012p250503}.
For LQR gates, the misorientation of the rotation axis occurs because the tunnel coupling has a DC component.
If all the DC components of the rotation axis are known, a three-step sequence could also be used to correct the misorientation in LQR.

\emph{Charging energy.}
When $g_0$ satisfies Eq.\ \eqref{eq:optimaldetst}, constraints on $\epsilon$ translate into constraints on $g_0$.
In the far-detuned regime, 
for the dot occupation to remain constant, $|\epsilon|$ must satisfy $|\epsilon|<U$, where $U$ is the
charging energy.
When $|\epsilon|\simeq U$, numerical optimization indicates that we may improve the fidelity slightly by deviating from Eq.\ \eqref{eq:optimaldetst}.

The results in Table~1 were obtained by numerically maximizing the fidelity.
The reported values of $f^\text{max}$ were obtained by using the most restrictive of the constraints described above.
In Table~1, we list the dominant constraints, and corresponding modifications that could enhance the fidelity.
Generally, we observe different constraints for different types of gates.
The STIRAP scheme appears particularly promising because optimization does not involve Eq.\ \eqref{eq:optimaldetst}.
(Hence, the charging energy constraint does not apply.)
Additional work is needed to clarify this scheme, however~\cite{footnote2}.
}

\begin{table}[t]
\centering 
\begin{tabular}
{@{\vrule height 10pt depth1pt  width0pt \extracolsep{\fill}}lcccc}
Qubit & $z$-gate & $x$: Pulsed & LQR & STIRAP~\cite{footnote2} \\
\hline
ST, Natural Si\tablenote{For the ST qubits we use $\Delta B=0.3$, 0.3, and 10~mT, respectively; for the hybrid qubits we use $\Delta E_{10}=0.1$~meV.}
& 97\tablenote{Constrained by the $z$-gate speed.  Improve by increasing $\Delta B$ or $\Delta E_{10}$.}
& 99.6\tablenote{Constrained by the charging energy ($U=1$~meV) and/or the misorientation of the rotation axis.  Improve by decreasing $\Delta B$ or increasing $U$.}
& 98$^\text{c}$\tablenote{Constrained by the RWA.  Improve by increasing $\Delta B$, $\Delta E_{10}$, or $U$.}
& 99.8\tablenote{Constrained by $g_0^\text{max}/h=10$~GHz.  Improve by increasing $g_0^\text{max}$.}
\\
ST, Purified $^{28}$Si$^\text{a}$
& 99.4$^\text{b}$
& 99.6$^\text{c}$
& 99.3$^\text{c}$$^\text{d}$
& 99.9$^\text{e}$
\\
ST, GaAs$^\text{a}$
& 66$^\text{b}$
& 98$^\text{c}$
& 92$^\text{c}$$^\text{d}$
& 98$^\text{e}$
\\
Hybrid, Natural Si$^\text{a}$
& 99.995$^\text{b}$
& 83\tablenote{Constrained because anticrossings must be distinct.  Improve by increasing $\Delta E_{10}$.}
& 99$^\text{c}$
& 99.6$^\text{e}$
\\
Hybrid, Purif. $^{28}$Si$^\text{a}$
& 99.999$^\text{b}$
& 83$^\text{f}$
& 99.4$^\text{c}$
& 99.8$^\text{e}$
\\
Hybrid, GaAs$^\text{a}$
& 94$^\text{b}$
& 83$^\text{f}$
& 89$^\text{c}$
& 96$^\text{e}$
\\
\hline
\end{tabular}
\caption{\bf Numerically maximized gate fidelities}  \label{table:1}
\end{table}

Next, we consider $z$-rotations.
Since $g_0$ (and hence $J$) can be turned off completely
and
the pure dephasing rate $\gamma$ is fixed, the gate fidelity can only optimized by 
maximizing the gate speed.
For a $\pi$-rotation, the gate period is $\tau=h/2\Delta E$, and the fidelity is
$f^\text{max}=(1+e^{-\gamma \tau})/2$, where $\Delta E=g\mu_B\Delta B$ for ST qubits and $\Delta E=\Delta E_{10}$ for hybrid qubits.
Here, we use $\Delta E_\text{nuc}=0.136$~neV for isotopically \suec{pure} $^{28}$Si (0.01\% $^{29}$Si), 3.0~neV for natural Si, and 92~neV for GaAs~\cite{Assali:2011p165301}.
Our results for $f^\text{max}$ are presented in Table~1.

The simulations reported here 
are for simple gating schemes,
including a one-step pulsed gate sequence for ST qubits, a five-step pulsed gate sequence for hybrid qubits, and a simple STIRAP scheme, which does not provide a true gate.
More sophisticated pulse sequences have also been proposed.
A three-step pulsed gate sequence was proposed to correct for the misorientation of the $x$-rotation axis in ST qubits~\cite{Hanson:2007p050502}.
{\color{black}
For pulsed gates, we find that this sequence does improve the fidelity over a small range of control parameters.  (See Appendix.)
However, the procedure incorporates an intermediate $z$-rotation step, so the final fidelity is bounded by the fidelity of the $z$-rotation.
Similar considerations should apply to LQR gates in ST qubits, although we do not study that problem here.
For hybrid qubits, the misorientation effect is quite weak for LQR, and corrective pulses have little effect.}

For $z$-rotations of ST qubits, because the noise spectrum of the nuclear spins is dominated by low frequencies~\cite{Neder:2011p035441}, pulse sequences similar to spin echoes~\cite{Bluhm:2011p109}
can 
improve the fidelity.
For hybrid qubits, the noise spectrum of optical phonons~\cite{Gamble:2012p035302} and charge fluctuators~\cite{Astafiev:2004p267007} has weight at higher frequencies, so echo-type pulse sequences may be less effective.
In principle, the fidelity of spin echoes can always be improved by increasing the sophistication of the pulse sequence~\cite{Lee:2008p160505}.
In practice, however, they are constrained by pulse imperfections and by
dephasing that occurs during the $x$-rotations in the sequence.

In conclusion, we have presented a method for optimizing the fidelity of gate operations of logical ST and hybrid qubits in the presence of both spin and charge dephasing, and we have identified upper bounds on the fidelity for simple gating schemes.
We obtain the following general results.
The fidelity of $z$-rotations in hybrid qubits in Si is high, because their gate speeds are much greater than their rate of pure dephasing.
The fidelity of $z$-rotations in ST qubits in dots without external field gradients is low but can be
improved greatly using spin echo methods.
Therefore, the limits on overall performance are those of the $x$-rotations.

For $x$-rotations in $^{28}$Si, the maximum achievable fidelities of
ST qubits (pulsed and LQR gates) and hybrid qubits (LQR gates) are similar, with $f^\text{max}> 99\%$.
STIRAP gates appear quite promising, although further work is required to clarify this scheme.
Hybrid qubits are probably not viable in GaAs due to 
\suec{fast}
pure dephasing~\cite{Gamble:2012p035302}.
For ST qubits, the maximum fidelities for $x$-rotations are considerably larger in Si than in GaAs. 
There are two reasons for this:  
(i) the large, intrinsic $\Delta B$ in GaAs causes a misorientation of the $x$-rotation axis, and 
(ii) the large $\gamma_\text{ST}$ in GaAs makes it difficult to implement corrective protocols involving $z$-rotations (\emph{e.g.}, \cite{Hanson:2007p050502}).
Hence, the nuclear spins that complicate the implementation of $z$-rotations in GaAs also ultimately constrain the $x$-rotations.

\section*{materials and methods}
The dynamical evolution of the logical qubit density matrix $\rho$ is governed by the master equation~\cite{Gardiner:2004}
{\small
\vspace{-.1in}
\begin{equation}\label{eq:master}
\frac{d\rho}{dt} = -\frac{i}{\hbar} [H, \rho] - D(\Gamma,\gamma),
\vspace{-.05in}
\end{equation}}
where $H$ is the Hamiltonian describing coherent evolution, and $D$ describes the fast ($\Gamma$) and slow ($\gamma$) dephasing processes.
While $D$ is phenomenological, its form can be justified in a bosonic environment, assuming Markovian dynamics, as has been argued for the double quantum dot system~\cite{Barrett:2002p125318}.
The analytical and numerical methods used
for solving Eq.\ \eqref{eq:master} are described in the Appendix.

To treat the ST and hybrid qubits on equal footing, we define our logical qubit basis states $|a\rangle$ and $|b\rangle$ in the far-detuned limit, $|\epsilon| \gg g_0$.  
(Note that for ST qubits this basis choice differs from that of
~\cite{Petta:2005p2180}.)
We also consider a third, excited charge state $|e\rangle$ that is tunnel-coupled to the logical qubit states for both the ST and hybrid qubit systems.
In the ${|a\rangle,|b\rangle,|e\rangle}$ ordered basis, we have
{\small \begin{equation} \label{eq:H}
H= 
\begin{pmatrix}
-\Delta E /2 & 0 & g_1 \\
 0 & \Delta E/2 & -g_2 \\
 g_1 & -g_2 & -\epsilon
\end{pmatrix},
\quad 
D = 
\begin{pmatrix}
0 & \gamma \rho_{ab} & \Gamma \rho_{ae} \\
\gamma \rho_{ba} & 0 & \Gamma \rho_{be} \\
\Gamma \rho_{ea} & \Gamma \rho_{eb} & 0 
\end{pmatrix}
\end{equation}}
In principle, the tunnel couplings $g_1$ and $g_2$ 
are independently tunable~\cite{Shi:2012p140503}; here
we take them to be equal, with $g=g_1=g_2$.

\begin{acknowledgments}
We thank Mark Eriksson, Jianjia Fei and Xuedong Hu for 
stimulating discussions\snc{,} and
acknowledge support from ARO (W911NF-12-1-0607), NSF (DMR-0805045, PHY-1104660), and the US DOD. The views and conclusions contained in this document are those of the authors and should not be interpreted as representing the official policies, either expressly or implied, of the U.S. Government.
\end{acknowledgments}


\begin{onecolumngrid}
\newcounter{subequation}
\renewcommand{\theequation}{S\arabic{subequation}}
\setcounter{subequation}{1}
\newcounter{subfigure}
\renewcommand{\thefigure}{S\arabic{subfigure}}
\setcounter{subfigure}{1}
\newcounter{subtable}
\renewcommand{\thetable}{S\arabic{subtable}}
\setcounter{subtable}{1}

\section*{Appendix:  Supplemental Information}

\section{Resonant gating:  Formalism}

\subsection{The interaction picture}
We consider a Hamiltonian $H$ 
{\color{black}that}
contains a time-independent component $H_0$ and a time-dependent part $V(t)$. 
In the Schr\"{o}dinger picture, the dynamics of the system are specified by the time-dependent wavefunction $|\psi(t)\rangle$, obtained by solving the Schr\"{o}dinger equation $i\hbar \partial_t |\psi(t)\rangle = H |\psi(t)\rangle$.
To study resonant phenomena, it is convenient to switch to the interaction picture~\cite{Tannor:2006}, defined by
\begin{equation}
|\psi(t)\rangle^= e^{-i H_0 t /\hbar} |\psi(t)\rangle^I,
\end{equation}  \addtocounter{subequation}{1}
where $|\psi(0)\rangle^I = |\psi(0)\rangle$. 
Here, we do not include any special labels to indicate Sch\"{o}dinger operators or states;
interaction quantities are indicated with the superscript $I$.
The interaction wavefunction obeys the equation of motion 
\begin{equation} \label{eq:psiI}
i \hbar \partial_t |\psi(t)\rangle^I = H^I(t) |\psi(t)\rangle^I,
\end{equation}  \addtocounter{subequation}{1}
where 
\begin{equation}
H^I(t) = e^{i H_0 t/\hbar} ~ V(t)~ e^{-i H_0 t/\hbar}. \label{eq:VI}
\end{equation}  \addtocounter{subequation}{1}
Expanding $|\psi(t)\rangle^I$ in an arbitrary basis $\{|n\rangle\}$,
\begin{equation}
|\psi(t)\rangle^I = \sum_n c_n(t) |n\rangle, \label{eq:cexpansion}
\end{equation}  \addtocounter{subequation}{1}
yields the equations of motion
\begin{equation}
i \hbar \dot{c}_m(t) = \sum_n V_{mn}(t) e^{i \omega_{mn} t} c_n(t).
\end{equation}  \addtocounter{subequation}{1}
Here, the matrix elements are defined as $V_{mn}(t) \equiv \langle m | V(t) | n\rangle$ and $\omega_{mn} \equiv \omega_m-\omega_n= (E_m - E_n)/\hbar$.

For both the ST and hybrid qubits, the resonant Hamiltonian involves the three-state basis, $\{ |a\rangle, |b\rangle, |e\rangle\}$, corresponding to the three-level systems shown in Fig.~1 of the main text. 
In this ordered basis, the Hamiltonian described in the main text is given by
\begin{equation} \label{eq:H}
H = \begin{pmatrix}
\hbar\omega_{a} & 0 & g_1(t) \\
0 & \hbar \omega_{b}  & -g_2(t) \\
g_1^*(t) & -g_2^*(t) & \hbar \omega_{e}
\end{pmatrix}, 
\end{equation}  \addtocounter{subequation}{1}
while the dephasing matrix is given by
\begin{equation} \label{eq:DST}
D = 
\begin{pmatrix}
0 & \gamma \rho_{ab} & \Gamma \rho_{ae} \\
\gamma \rho_{ba} & 0 & \Gamma \rho_{be} \\
\Gamma \rho_{ea} & \Gamma \rho_{eb} & 0 
\end{pmatrix} .
\end{equation}  \addtocounter{subequation}{1}
Dividing up the terms in Eq.~(\ref{eq:H}), we arrive at the following definitions for the interaction picture:
\begin{equation} \label{eq:intdef}
H_0 = \left( \begin{array}{ccc}
\hbar\omega_{a} & 0 & 0 \\
0 &\hbar \omega_{b} & 0 \\
0 & 0 & 0
\end{array} \right),~
V(t) = \left( \begin{array}{ccc}
0 & 0 & g_1(t) \\
0 & 0 & -g_2(t) \\
g_1^*(t) & -g_2^*(t) & \hbar \omega_{e}
\end{array} \right), 
\end{equation}  \addtocounter{subequation}{1}
yielding
\begin{equation}
H^I(t) = \begin{pmatrix}
0 & 0 & g_1(t)e^{i\omega_at} \\
0 & 0 & -g_2(t)e^{i\omega_bt} \\
g_1^*(t)e^{-i\omega_at} & -g_2^*(t)e^{-i\omega_bt} & \hbar \omega_{e}
\end{pmatrix} . \label{eq:HI}
\end{equation}  \addtocounter{subequation}{1}
Here, we have chosen to include $\hbar \omega_e$ in $V(t)$ as a matter of convenience, although it is not time-dependent.
For both types of qubits, we define the energy zero to be half way between $\hbar\omega_a$ and $\hbar\omega_b$, as indicated in Fig.~1(a) of the main text.
Similarly, $\hbar \omega_e=-\epsilon$, while $\hbar \omega_{a,b}=\mp \Delta E_B/2$ for ST qubits, and 
$\hbar \omega_{a,b}=\mp \Delta E_{10}/2$ for hybrid qubits.
We have included indices on the tunnel couplings $g_{1,2}(t)$ for completeness, because they should be independently tunable in hybrid qubits~\cite{Shi:2012p140503}.
However, in the following analysis, we assume they are equal, with $g(t)=g_1(t)=g_2(t)$.
Resonance is achieved by modulating the tunnel coupling with a tunable driving frequency $\omega$.
As discussed in the main text, we assume the sign of the tunnel coupling remains fixed, with the form
\begin{equation}
g(t)= g_0 \left[1 + \mathrm{cos}\left(\omega t \right) \right]. \label{eq:g2}
\end{equation}  \addtocounter{subequation}{1}
Note that this expression includes both AC and DC components.

\subsection{Derivation of the conditions for resonance}
It is helpful to solve the time evolution of the three-level system, to identify the resonances that emerge.
In general, the problem cannot be solved exactly. 
However, it is possible to 
identify resonant terms
by employing time-dependent perturbation theory.

We expand the basis coefficients in Eq.~(\ref{eq:cexpansion}) in powers of the interaction:
\begin{equation}
c_n(t) = c_n^{(0)} + c_n^{(1)}(t) + c_n^{(2)}(t) + \dots
\end{equation}  \addtocounter{subequation}{1}
where $c_n^{(i)} \sim \mathcal{O}(V^i)$ and $c_n^{(0)}$ represents the initial state at time $t=0$. 
The next two terms  in the expansion are
\begin{eqnarray}
c_n^{(1)}(t) &=& \!-\frac{i}{\hbar} \int_{t_0}^{t} dt' e^{i \omega_{ni} t'} V_{ni}(t') \label{eq:c1} ,\\
\addtocounter{subequation}{1}
c_n^{(2)}(t) &=& \!-\frac{1}{\hbar^2} \int_{t_0}^{t} \! dt' \int_{t_0}^{t'} \! dt{''} e^{i \omega_{nm} t' + i \omega_{mi} t''} V_{nm}(t') V_{mi}(t''). \label{eq:c2} 
\end{eqnarray} \addtocounter{subequation}{1}
The probability of transitioning from the initial state $|i\rangle$ to state $|n\rangle$ (for $n\neq i$) is then given by $P_{i\to n}(t) =\left |c_n(t) \right |^2 = \left | c_n^{(0)} + c_n^{(1)}(t) + c_n^{(2)}(t) + \dots \right |^2$.

We calculate to second order in the couplings, so 
for initial state $|a\rangle$, we approximate the probability of transitioning to the final state $|b\rangle$ as
\begin{equation}
P_{a\to b}(t) \simeq \left| c_b^{(0)} + c_b^{(1)}(t) + c_b^{(2)}(t) \right|^2. \label{eq:P01}
\end{equation}  \addtocounter{subequation}{1}
For the initial state $c_a^{(0)}=1$, $c_b^{(0)}=0$, the first term on the right hand side of Eq.~(\ref{eq:P01}) vanishes. 
The second term, $c_b^{(1)}(t) $, also vanishes because $V_{ba} = 0$. 
The third term, which is second order in the perturbation expansion, is given by
\begin{eqnarray}
c_b^{(2)}(t) &=& -\frac{1}{\hbar^2} \int_{0}^{t} dt' \int_{0}^{t'} dt{''} e^{-i \omega_{eb} t' + i \omega_{ea} t''} V_{be}(t') V_{ea}(t'') \nonumber \\
&=&  \frac{g_0^2}{\hbar^2} \int_{0}^{t} dt' \int_{0}^{t'} dt{''} e^{-i \omega_{eb} t' + i \omega_{ea} t''} \left[1+\mathrm{cos}(\omega t')\right]  \left[1+\mathrm{cos}(\omega t'')\right]  
=\beta_1+\beta_2+\beta_3+\beta_4 , \label{eq:c12}
\end{eqnarray}\addtocounter{subequation}{1}
where
\begin{eqnarray}
\beta_{1} &\equiv& \frac{g_0^2}{\hbar^2} \int_{0}^{t} dt' \int_{0}^{t'} dt{''} e^{-i \omega_{eb} t' + i \omega_{ea} t''} , \\ \addtocounter{subequation}{1} 
\beta_{2}&\equiv& \frac{g_0^2}{\hbar^2} \int_{0}^{t} dt' \int_{0}^{t'} dt{''} e^{-i \omega_{eb} t' + i \omega_{ea} t''} \mathrm{cos}(\omega t') \mathrm{cos}(\omega t'') , \label{eq:b2a} \\ \addtocounter{subequation}{1} 
\beta_{\text{3}}&\equiv& \frac{g_0^2}{\hbar^2} \int_{0}^{t} dt' \int_{0}^{t'} dt{''} e^{-i \omega_{eb} t' + i \omega_{ea} t''} \mathrm{cos}(\omega t'') , \\ \addtocounter{subequation}{1} 
\beta_{\text{4}}&\equiv& \frac{g_0^2}{\hbar^2} \int_{0}^{t} dt' \int_{0}^{t'} dt{''} e^{-i \omega_{eb} t' + i \omega_{ea} t''} \mathrm{cos}(\omega t') . 
\end{eqnarray} \addtocounter{subequation}{1} 
Integration yields
\begin{eqnarray}
\beta_{1} &=& \frac{g_0^2}{\hbar^2} \left(\frac{i}{\omega_{ea}}\right)\left( -\frac{1-e^{-i \omega_{eb} t}}{\omega_{eb}} + \frac{1-e^{i\omega_{ba}t}}{\omega_{ba}}\right), \label{eq:b1} \\ \addtocounter{subequation}{1} 
\beta_{2} &=& \frac{ g_0^2}{4\hbar^2}\left[ \frac{1}{\omega_{ea}+\omega}\left( \frac{1-e^{i(-\omega_{eb}+\omega)t}}{-\omega_{eb}+\omega} +\frac{1-e^{i(-\omega_{eb}-\omega)t}}{-\omega_{eb}-\omega}  -\frac{1-e^{i(\omega_{ba}+2\omega)t}}{\omega_{ba}+2\omega} -\frac{1-e^{i\omega_{ba}t}}{\omega_{ba}} \right) \right. \nonumber \\ 
&& ~~~~~~\left. +  \frac{1}{\omega_{ea}-\omega}\left(\frac{1-e^{i(-\omega_{eb}+\omega)t}}{-\omega_{eb}+\omega} +\frac{1-e^{i(-\omega_{eb}-\omega)t}}{-\omega_{eb}-\omega}  -\frac{1-e^{i(\omega_{ba}-2\omega)t}}{\omega_{ba}-2\omega} -\frac{1-e^{i\omega_{ba}t}}{\omega_{ba}}  \right) \right], \label{eq:b2} \\ \addtocounter{subequation}{1} 
\beta_{\text{3}}&=& \frac{ g_0^2}{2\hbar^2}\left[ \frac{1}{\omega_{ea}+\omega} \left( -\frac{1-e^{-i\omega_{eb}t}}{\omega_{eb}} -\frac{1-e^{i(\omega_{ba}+\omega)t}}{\omega_{ba}+\omega}  \right) 
+ \frac{1}{\omega_{ea}-\omega}\left( -\frac{1-e^{-i\omega_{eb}t}}{\omega_{eb}}   -\frac{1-e^{i(\omega_{ba}-\omega)t}}{\omega_{ba}-\omega}  \right) \right], \\ \addtocounter{subequation}{1} 
\beta_{\text{4}}&=&\frac{g_0^2}{2\hbar^2}\frac{1}{\omega_{ea}} \left(  \frac{1-e^{i(-\omega_{eb}+\omega)t}}{-\omega_{eb}+\omega}  +  \frac{1-e^{i(-\omega_{eb}-\omega)t}}{-\omega_{eb}-\omega} - \frac{1-e^{i(\omega_{ba}+\omega)t}}{\omega_{ba}+\omega}  - \frac{1-e^{i(\omega_{ba}-\omega)t}}{\omega_{ba}-\omega}  \right) . \label{eq:b4}
\end{eqnarray} \addtocounter{subequation}{1} 

Many of the individual terms appearing in Eqs.~(\ref{eq:b1})-(\ref{eq:b4}) are rapidly oscillating and small, (of order $g_0^2/\hbar^2\omega_{ea}\omega_{ba}$, or smaller).
However, $P_{a\rightarrow b}$ 
is strongly peaked at special resonant values of $\omega$.
Several resonances can be identified.  
The expression for $\beta_2$ contains the conventional resonant terms, arising from the purely AC modulation of the tunnel coupling, corresponding to the $\cos (\omega t')\cos (\omega t'')$ term in Eq.~(\ref{eq:b2a}). 
In Eq.~(\ref{eq:b2}), we see that this ``primary" resonance occurs when $\omega =\omega_{ba}/2$.
In conventional ESR, the primary resonance would occur at the frequency $\omega_{ba}$.
For LQR, it occurs at $\omega_{ba}/2$ because the process is not directly between the states $|a\rangle$ and $|b\rangle$; rather, it is mediated by the excited state $|e\rangle$.
The resulting $\beta_2$ resonance terms are resonant when $2\omega = \omega_{ba}$.
In $\beta_2$, we also observe resonances at the excitation frequencies $\omega_{ea}$ and $\omega_{eb}$, which are used to drive the STIRAP protocol, discussed below.
The tunnel coupling in Eq.~(\ref{eq:g2}) also contains a DC component.
The purely DC contribution to $P_{a\rightarrow b}$ is contained in $\beta_1$, and it causes no resonant excitations, as expected.  
The AC-DC cross terms in $P_{a\rightarrow b}$ are found in $\beta_3$ and $\beta_4$.
Here, we observe a new secondary resonance occuring at $\omega=\omega_{ba}$.

In the limit of weak coupling ($g_0 \ll \hbar \omega_{ea}$) and large detuning ($\omega_{ba} \ll \omega_{ea}$), we can identify the leading terms in the transition probability arising from the primary and secondary resonances:
\begin{eqnarray}
\lim_{\omega \to \omega_{ba}/2} P_{a\to b} (t)&\simeq & \frac{g_0^4}{16\hbar^4  \omega_{ea}^2} ~\left\{ \frac{ \sin \left[ (\omega-\omega_{ba}/2)t \right] }{(\omega-\omega_{ba}/2) } \right\}^2\label{eq:resonance10}, \\ \addtocounter{subequation}{1} 
\lim_{\omega \to \omega_{ba}} P_{a\to b} (t)&\simeq& \frac{g_0^4}{\hbar^4 \omega_{ea}^2} ~\left\{ \frac{ \sin \left[ (\omega-\omega_{ba})t/2 \right] }{(\omega-\omega_{ba})/2 } \right\}^2 \label{eq:resonancehalf}.
\end{eqnarray} \addtocounter{subequation}{1} 
In the long-time limit ($t\gg (\Delta \omega)^{-1}\sqrt{3/2}$), the resonance function $\left\{ \mathrm{sin}(\Delta \omega\, t) / \Delta \omega \right\}^2$ is peaked at $\Delta \omega= 0$, with a height of $t^2$, and a full-width-at-half-max of $\sqrt{6}/t$.  
Similar results can be obtained for the $\omega_{ea}$ and $\omega_{eb}$ resonances.

\subsection{Rotating wave approximation (RWA)}
When performing simulations involving fast driving frequencies, it is convenient to explicitly account for resonant effects by applying the rotating wave approximation (RWA) \cite{Shore:1990}.  
The idea is to drop the small fast-oscillating terms in the time evolution, which average to zero, while retaining the large constant terms.
For example, at the primary resonance, we can compute the time-averaged component of Eq.~(\ref{eq:HI}), by plugging in Eq.~(\ref{eq:g2}).
The calculation is more straightforward when we use $-\hbar \omega_a=\hbar \omega_b=\hbar \omega_{ba}/2=\Delta E/2$, yielding the time-independent form
\begin{equation}
H^{I} \simeq \begin{pmatrix}
0 & 0 & \tilde{g} \\
0 & 0 & -\tilde{g} \\
\tilde{g} & -\tilde{g} & \hbar \omega_e
\end{pmatrix},  \label{eq:Hrwa1half} 
\end{equation}  \addtocounter{subequation}{1}
where $\tilde{g}=g_0/2$.
We have confirmed the validity of this approximation by comparing numerical time evolutions obtained using the exact form, Eq.~(\ref{eq:HI}), and the approximate form, Eq.~(\ref{eq:Hrwa1half}).

The RWA approximation should be accurate if many cycles of resonant oscillations occur inside a single LQR pulse.
This requirement can be quantified, as follows.
The gate time for an LQR $\pi$-pulse is given by $\tau_\pi \simeq \pi\hbar^2\omega_e/2\tilde{g}^2$, while the time for a single resonant oscillation is $\tau_r=4\pi/\omega_{ba}$.
The requirement that $\tau_r \ll \tau_\pi$ can then be expressed as  
$g_0\ll \sqrt{(\hbar\omega_e)(\hbar\omega_{ba})/2}$ for the primary resonance.

The secondary resonance $\omega=\omega_{ab}$ also yields an approximately time-independent interaction Hamiltonian.
Numerical investigation indicates that the effective $H^I$ takes the same form as Eq.~(\ref{eq:Hrwa1half}), except with $\tilde{g}=g_0$.
In this case, $\tau_r=2\pi/\omega_{ba}$, so the validity requirement for the RWA becomes $g_0\ll \sqrt{(\hbar\omega_e)(\hbar\omega_{ba})/4}$ for the secondary resonance.
All numerical simulations of LQR reported in this work use the time-independent form for $H^I$, given in Eq.~(\ref{eq:Hrwa1half}).

\section{Analysis of LQR}
In this section we focus on the LQR resonant gate.
Our goal is to calculate the fidelity of a $\pi$-rotation around the $x$-axis.
We first discuss the definition of fidelity used in this work.
We then describe the analytical estimates of the fidelity reported in the main text and the procedure used to optimize the fidelity.

Based on experimental observations in semiconducting qubits~\cite{Petta:2005p2180, Bluhm:2011p109, Petersson:2010p246804, Shi:2012preprint}, the results reported in the main
text incorporate two different dephasing rates reflecting different dephasing mechanisms.
For small detuning values, the fast dephasing ($\Gamma$) due to charge noise dominates
because of the mixing with the excited state with different charge character, while for large detuning, the slow (pure) dephasing ($\gamma$) associated with the spin sector dominates. 
Below, we treat these two regimes separately.

\subsection{Definition of fidelity}\label{sec:fidelity}
{\color{black}We define the state fidelity $f$ in terms of the actual density matrix $\rho$ (obtained by solving a master equation) and the target state $|\phi\rangle$:
\begin{equation}\label{eq:densitymatrixdist}
f  = \langle\phi | \rho |\phi\rangle.
\end{equation}  \addtocounter{subequation}{1}}
The results presented in the main text  for $x$-rotations are all for  $\pi$-rotations 
beginning with the initial state $\rho(0)=|a\rangle\langle a|$. 
(The trends for other rotation angles and other initial states are similar.)
We use the master equation described below to compute the actual density matrix $\rho(\tau_\pi)$.
In this case, a perfect $\pi$-rotation in a gate period $\tau_\pi$
yields the target state $|b\rangle$.
The state fidelity is therefore given by
\begin{equation}
f = \langle b|\rho(\tau_\pi) |b\rangle = \rho_{bb}(\tau_\pi) .
\end{equation}  \addtocounter{subequation}{1}
The fidelity for a perfect gate operation is $f=1$.
Below, we compute the optimal fidelity $f^\text{opt}$, as well as the infidelity $1-f^\text{opt}$.

\subsection{Incorporation of dephasing into the calculation}
As described above, we solve for the density matrix in the interaction picture, defined as
\begin{equation} 
\rho^{I} (t) = e^{iH_0 t/\hbar} \rho(t) e^{-iH_0 t/\hbar} .
\end{equation}  \addtocounter{subequation}{1}
The master equation in the interaction picture is given by~\cite{Gardiner:2004}
\begin{equation}\label{eq:masterintpic}
\dot{\rho}^{I} (t) = -\frac{i}{\hbar} [ H^{I}(t) , \rho^{I}(t) ] - D^{I},
\end{equation}  \addtocounter{subequation}{1}
where $H^I$ is the time-independent Hamiltonian obtained in the RWA, Eq.~(\ref{eq:Hrwa1half}), and $D^I= e^{iH_0 t/\hbar} D e^{-iH_0 t/\hbar}$ captures the dephasing effects.
The dephasing matrix given in Eq.~(\ref{eq:DST}) describes double-occupation errors occurring during the exchange interaction~\cite{Barrett:2002p125318}, which can be understood as charge noise associated with the tunnel coupling $g$.
The noise considered in \cite{Barrett:2002p125318} is strictly Markovian.
As argued in the main text, our focus on $g$-noise rather than $\epsilon$-noise is appropriate in the far-detuned regime, which is the regime of interest because it is 
where the maximum fidelity $f^\text{max}$ is largest.
In the far-detuned regime, the effects of  $\epsilon$-noise are
 suppressed because the energies of the
qubit states depend very similarly on detuning.  
For the $D$ matrix in Eq.~(\ref{eq:DST}), which incorporates dephasing with
fast decay rate $\Gamma$
(between the excited charge state and each qubit state) and dephasing with
slow decay rate $\gamma$ (between
the two qubit states), we obtain
\begin{equation} \label{eq:DI}
D^I = 
\begin{pmatrix}
0 & \gamma \rho^I_{ab} & \Gamma \rho^I_{ae} \\
\gamma \rho^I_{ba} & 0 & \Gamma \rho^I_{be} \\
\Gamma \rho^I_{ea} & \Gamma \rho^I_{eb} & 0 
\end{pmatrix} .
\end{equation}  \addtocounter{subequation}{1}

\subsection{Analytic approximation for the fidelity in LQR}
The results presented for LQR in the main text are numerical solutions of Eqs.~(\ref{eq:masterintpic}) and (\ref{eq:DI}).
We assume that the tunnel coupling pulse in Eq.~(\ref{eq:g2}) is turned on and off suddenly,
taking $g_0$ to be a step function pulse envelope: 
\begin{equation} \label{eq:gstep}
g_0(t)=\left\{ \begin{array}{ccc}
g_0 & & (0\leq t\leq \tau_\pi) \\
0 & & \text{(otherwise)} 
\end{array} \right. .
\end{equation}  \addtocounter{subequation}{1}
Fig.~1(d) of the main text shows a softer pulse envelope, which is more physically realistic; however for simplicity, we consider Eq.~(\ref{eq:gstep}) here.
This approximation is reasonable because we are focusing on the contribution at the
resonant frequency using the RWA.
(We consider smooth, Gaussian pulse envelopes for the STIRAP calculations described below.)

In the next two subsections, we obtain analytical estimates for the fidelity in the regimes
where fast dephasing and slow dephasing dominate.

\subsubsection{Fast dephasing: $\Gamma$-dominant regime}
We first consider the hybrid qubit in the regime where fast dephasing dominates.
In this regime, we can approximate
$\gamma \rightarrow 0$ in Eq.~(\ref{eq:DI}), which allows us to make analytical progress.
We will show later that these results are in excellent agreement with exact numerical results.

We first perform a change of basis:
\begin{gather}
|u\rangle = \frac{1}{\sqrt{2}}\left(|a\rangle - |b\rangle \right),\\ \addtocounter{subequation}{1} 
|v\rangle = \frac{1}{\sqrt{2}}\left(|a\rangle + |b\rangle\right),\\ \addtocounter{subequation}{1} 
|w\rangle = |e\rangle.
\end{gather} \addtocounter{subequation}{1} 
In this basis, Eq.~(\ref{eq:densitymatrixdist}) can be rewritten as
\begin{eqnarray} \label{eq:f2}
f = \frac{1}{2}\left[ \rho_{uu}^I (\tau_\pi) + \rho_{vv}^I (\tau_\pi) - \rho_{uv}^I (\tau_\pi) - \rho_{vu}^I (\tau_\pi) \right].
\label{eq:fhy2}
\end{eqnarray} \addtocounter{subequation}{1} 
Henceforth, unless otherwise noted, we drop the $I$ superscript to simplify the notation.

Equations~(\ref{eq:masterintpic}) and (\ref{eq:DI}) yield the following set of six independent equations:
\begin{eqnarray}
\dot{\rho}_{vv}  &=& 0,\label{eq:rhovv}\\ \addtocounter{subequation}{1} 
\dot{\rho}_{vu}  &=& \frac{i}{\hbar}  \sqrt{2}\,\tilde{g} \, \rho_{vw} ,\label{eq:rhovu}\\ \addtocounter{subequation}{1} 
\dot{\rho}_{vw}  &=&  \frac{i}{\hbar} \left[  \sqrt{2}\,\tilde{g} \,  \rho_{vu}  + |\epsilon| \rho_{vw}  \right]- \Gamma \rho_{vw} ,\label{eq:rhovw}\\ \addtocounter{subequation}{1} 
\dot{\rho}_{uu}  &=& \frac{i}{\hbar}  \sqrt{2}\,\tilde{g} \,  (\rho_{uw}  - \rho_{uw}^*),\label{eq:rhouu}\\ \addtocounter{subequation}{1} 
\dot{\rho}_{uw}  &=& \frac{i}{\hbar} \left[ \sqrt{2}\,\tilde{g} \, (\rho_{uu} -\rho_{ww} ) +| \epsilon |\rho_{uw} \right]- \Gamma \rho_{uw} ,\label{eq:rhouw}\\ \addtocounter{subequation}{1} 
\dot{\rho}_{ww}  &=& -\frac{i}{\hbar} \sqrt{2}\,\tilde{g} \,  ( \rho_{uw}  - \rho_{uw}^*).\label{eq:rhoww}
\end{eqnarray} \addtocounter{subequation}{1} 
Here, we make the replacement $\hbar \omega_e=-\epsilon$, as consistent with the main text, including Fig.~1.
Since $\epsilon \leq 0$ in this work, we will often use an absolute value sign to avoid confusion ($|\epsilon|$).

Equations~(\ref{eq:rhovv})-(\ref{eq:rhoww}) can be separated into three decoupled blocks.
Equation~(\ref{eq:rhovv}) is trivially decoupled from the others.
Equations~(\ref{eq:rhovu}) and (\ref{eq:rhovw}) form a two-fold coupled block, while Eqs.~(\ref{eq:rhouu})-(\ref{eq:rhoww}) form a three-fold coupled block. 
Within a given block, the partial density (\emph{i.e.}, the sum of the diagonal terms in $\rho$) is conserved.
We now solve the separate blocks of equations analytically.
For the fidelity calculation specified in Eq.~(\ref{eq:f2}), the initial conditions are given by $\rho_{uu}(0) = \rho_{vv}(0) = \rho_{vu}(0) = 1/2$, and $\rho_{ww}(0)=\rho_{vw}(0)=\rho_{uw}(0)=0$.

The fully decoupled Eq.~(\ref{eq:rhovv})  trivially yields the result
\begin{equation}
\rho_{vv} (t) = \rho_{vv} (0)=\frac{1}{2}. \label{eq:sol2}
\end{equation}  \addtocounter{subequation}{1}

The two-fold block of equations (\ref{eq:rhovu}) and (\ref{eq:rhovw}) can be solved, yielding an exact solution:
\begin{equation} \label{eq:rhovu2}
\rho_{vu} (t)= \frac{1}{4 \xi} 
\left[ \left(\xi -\Gamma + \frac{i |\epsilon|}{\hbar} \right) +e^{\xi  t} \left(\xi+\Gamma -  \frac{i |\epsilon|}{\hbar} \right) \right] e^{-\frac{1}{2}t(\xi + \Gamma -  i|\epsilon|/\hbar )} ,
\end{equation}  \addtocounter{subequation}{1} 
where 
\begin{equation}
\xi = \sqrt{ \left(\Gamma - \frac{i |\epsilon|}{\hbar} \right)^2 -8\left(\frac{\tilde{g}}{\hbar}\right)^2 } .\label{eq:eta}
\end{equation}  \addtocounter{subequation}{1}
Working to leading order in the weak-tunneling, slow-dephasing limit, $\tilde{g},\Gamma \ll |\epsilon|$, we obtain
\begin{equation}\label{eq:sol3}
\rho_{vu} (t) \simeq \frac{1}{2} e^{- ( \tilde{\Gamma} + i J/\hbar )t} ,
\end{equation}  \addtocounter{subequation}{1}
where
\begin{equation}
\tilde{\Gamma} = \Gamma\frac{2\tilde{g}^2}{\epsilon^2} \quad\quad \text{and} \quad\quad
J  = \frac{2\tilde{g}^2}{|\epsilon|}. 
\end{equation}  \addtocounter{subequation}{1}

The three-fold block of equations (\ref{eq:rhouu})-(\ref{eq:rhoww}) can also be solved exactly, following the method of Ref.~\cite{Barrett:2002p125318}.
We parameterize the density matrix elements $\rho_{uu}$, $ \rho_{uw}$, and $\rho_{ww}$ in terms of a state vector in the two-dimensional $\{ |u\rangle, |w\rangle \}$ manifold.
We have noted that the $\{ |u\rangle, |w\rangle \}$ manifold is closed; however, the radius of the corresponding Bloch sphere is less than 1 since the density of the hybrid qubit does not lie entirely within the manifold.  
To remedy this, we can make use of the conserved quantity $[\rho_{uu}(t)+\rho_{ww}(t)]$ to define
\begin{equation}\label{eq:Blochuw}
\begin{pmatrix}
\rho_{uu} & \rho_{uw}  \\
\rho_{uw}^* & \rho_{ww}
\end{pmatrix}
=[\rho_{uu}(0)+\rho_{ww}(0)]\frac{\left( I + {\bf n} \cdot \bm{\sigma} \right)}{2}.
\end{equation}  \addtocounter{subequation}{1}
Here, $\bf n$ represents a state in the $\{ |u\rangle, |w\rangle \}$ manifold on a \emph{unit} Bloch sphere.
From Eqs.~(\ref{eq:rhouu})-(\ref{eq:rhoww}), we then obtain a compact description of the damped precession of the Bloch vector about an effective magnetic field:
\begin{equation}\label{eq:ndot}
\dot{{\bf n}} = \bm{\Omega} \times {\bf n} - \Gamma {\bf n}_{t} .
\end{equation}  \addtocounter{subequation}{1}
Here, ${\bf n}_{t} \equiv (n_x, n_y, 0)$ is the transverse component of the Bloch vector, and the effective magnetic  field  is given by $\bm{\Omega} = (2\sqrt{2}\,\tilde{g}/\hbar ,0, |\epsilon|/\hbar)$. 
Since $\tilde{g}\ll |\epsilon|$, the effective field makes a very small angle with the $z$ axis.
The first term on the right-hand-side of Eq.~(\ref{eq:ndot}) describes the coherent precession of the Bloch vector about the effective field. 
The second term describes the damping of the transverse component of the Bloch vector.  
The initial state corresponds to ${\bf n}_{0}=(0,0,1)$.

If the gate operation is implemented adiabatically and the dephasing rate is low enough
to satisfy $h \Gamma \ll |\epsilon|$, the Bloch vector precesses around $\bm{\Omega}$ with a small angle during the entire time evolution~\cite{Barrett:2002p125318}; in this small-angle limit we can obtain an analytical solution to Eq.~(\ref{eq:ndot}).
Taking the dot product with $\bm{\Omega}$ on both sides of Eq.~(\ref{eq:ndot}) and applying the small-angle approximation, the equation simplifies to
\begin{equation}\label{eq:lengthn}
 \dot{n} \simeq -\Gamma ~\mathrm{sin}^2 \theta ~n \simeq -\left(\frac{8\tilde{g}^2}{\epsilon^2}\Gamma\right) n,
 \end{equation}  \addtocounter{subequation}{1}
where $n$ is the length of the Bloch vector and $\theta \simeq \mathrm{arcsin}\left(8\tilde{g}^2/\epsilon^2 \right)$ is the angle between the effective field $\bm{\Omega}$ and the $z$-axis of the Bloch sphere. 
Averaged over many precession cycles, the Bloch vector $\bf n$ makes an angle of approximately $\theta$ with the $z$-axis. 
Since angle $\theta$  is small, we can approximate $n_z \simeq n$. 
Solving Eq.~(\ref{eq:lengthn}) in the small-angle approximation for $\theta$, we obtain
\begin{equation}
n_z(t) \simeq \mathrm{exp} \left( - \frac{8 \tilde{g}^2}{\epsilon^2} \Gamma t\right).
\end{equation}  \addtocounter{subequation}{1} 
Equation~(\ref{eq:Blochuw}) then yields
\begin{equation}
\rho_{uu}(t) = \frac{1}{4} \left[1 + n_z(t)\right]
\simeq \frac{1}{4} \left[1+ \mathrm{exp}\left(-\frac{8\tilde{g}^2}{\epsilon^2}
 \Gamma t \right) \right ].\label{eq:sol1}
\end{equation}  \addtocounter{subequation}{1}
Finally, from Eq.~(\ref{eq:f2}) we obtain
\begin{equation}\label{eq:fchy}
 f^\text{f} \simeq \frac{3}{8}+\frac{1}{8}e^{- 4 \pi \hbar\Gamma/|\epsilon| } + \frac{1}{2}e^{-\pi \hbar \Gamma/|\epsilon|}.
\end{equation}  \addtocounter{subequation}{1}
where we have used $\tau_\pi = \pi\hbar /J$.
Here, the superscript `f' refers to the fast dephasing regime.
For high fidelities, Eq.~(\ref{eq:fchy}) can be rewritten as
\begin{equation}
\label{eq:fchy2}
 f^\text{f} \simeq 1-h\Gamma/2|\epsilon|.
\end{equation}  \addtocounter{subequation}{1}

Some typical results for the infidelity are shown in Fig.~\ref{fig:Vdetuning}, for both ST and hybrid qubits.
We see that the form derived in Eq.~(\ref{eq:fchy}) approximates the numerical results very well at larger values of $|\epsilon|^{-1}$.
In this regime, the fidelity of the LQR gate is nearly independent of the tunnel coupling; it depends only on the ratio $\hbar \Gamma / |\epsilon|$. 
At smaller values of $|\epsilon|^{-1}$, Eq.~(\ref{eq:fchy}) is inaccurate because pure
dephasing %
with decay rate $\gamma\ne 0$ dominates,
as discussed in the following subsection.

\begin{figure}
\begin{center}
\centerline{\includegraphics[scale=0.38]{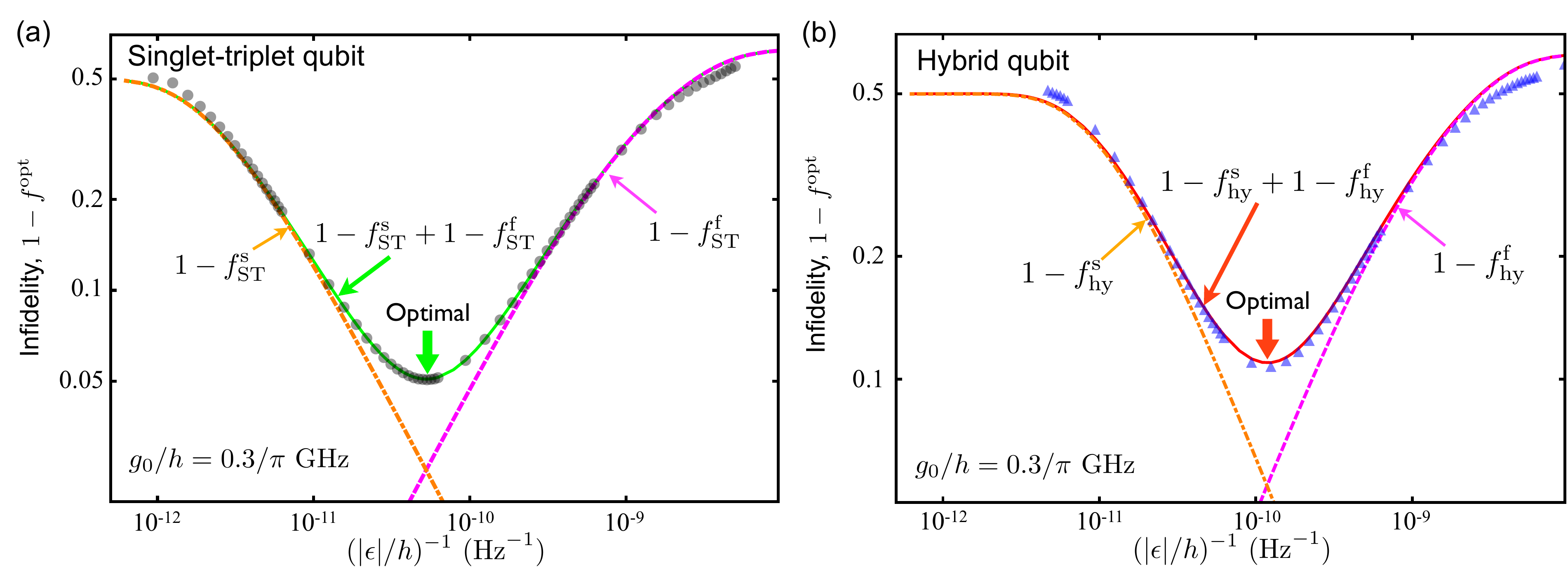}}
\caption{\label{fig:Vdetuning}
Plots of the infidelity (1-fidelity) vs.\ inverse detuning $(|\epsilon|/\hbar)^{-1}$ for an LQR gate, while keeping the tunnel coupling fixed with $g_0/h = 0.3/\pi~\mathrm{GHz}$.
(a) ST qubit.
(b) Hybrid qubit.
The same numerical results (indicated by markers) were presented in the inset of Fig.~2 in the main text. 
The dashed, magenta curves on the right-hand-side of the plots correspond to the ``fast" dephasing-dominated behavior, $f^\text{f}$, given in Eq.~(\ref{eq:fchy}).
The dashed, orange curves on the left-hand-side of the plots correspond to the ``slow" or ``pure"
dephasing-dominated behavior, $f^\text{s}$, given in Eq.~(\ref{eq:fshy}).
The solid curves represent the sums of the fast and slow infidelity curves.
This simple approximation matches the numerical results quite well, and the intersection of the fast and slow infidelity curves provides a good approximation for the optimal detuning $\epsilon^\text{opt}$.
}
\end{center}
\end{figure}
\addtocounter{subfigure}{1}  

\subsubsection{Slow dephasing: $\gamma$-dominant regime}
We now consider the regime where slow, pure dephasing dominates.
In this limit, we make the approximation $\Gamma = 0$ in Eq.~(\ref{eq:DI}).
Since charge noise is ignored, and since the tunnel coupling to the excited state $|e\rangle$ is weak, we may apply a Schrieffer-Wolff transformation~\cite{Schrieffer:1966p491} to eliminate $|e\rangle$.
In the remaining two-state system (\emph{i.e.}, $\{|a\rangle,|b\rangle\}$), the second-order tunneling process is replaced by an effective exchange coupling, $J =2 \tilde{g}^2/|\epsilon|$.
The effective interaction Hamiltonian is given by
\begin{equation}
H^{I ,\text{eff}} =\frac{1}{2}
\begin{pmatrix}
0 & J \\
J & 0
\end{pmatrix} .
\end{equation}  \addtocounter{subequation}{1}
In the two-state system, the master equation is still given by Eq.~(\ref{eq:masterintpic}), while the dephasing matrix is given by
\begin{equation}
D^{I,\text{eff}} =
\begin{pmatrix}
0 & \gamma \rho_{ab}^I \\
\gamma \rho_{ba}^I & 0
\end{pmatrix} .
\end{equation}  \addtocounter{subequation}{1}

The master equation yields two coupled differential equations for the density matrix elements.
Again dropping the $I$ superscript, we have
\begin{eqnarray}
\dot{\rho}_{bb} &=& -\frac{i J}{2 \hbar} \left( \rho_{ba}^* - \rho_{ba} \right),\\ \addtocounter{subequation}{1} 
\dot{\rho}_{ba} &=& -\frac{i J}{2 \hbar} \left( 1-2 \rho_{bb} \right)  - \gamma \rho_{ba} .
\end{eqnarray}  \addtocounter{subequation}{1} 
These yield an approximate solution which is accurate up to first order in $\hbar \gamma/J$, given by
\begin{equation}
\rho_{bb}(t) \simeq \frac{1}{2}\left[ 1 -e^{-\gamma t/2} \cos ( Jt/\hbar) \right],
\end{equation}  \addtocounter{subequation}{1}
The fidelity for a $\pi$-rotation is then given by
\begin{equation}
 f^{\text{s}} =  \rho_{bb} (\tau_\pi)  =  \rho_{bb}^I(\tau_\pi)  \simeq
 \frac{1}{2} \left[1 + \mathrm{exp}\left(-\frac{\pi \hbar \gamma |\epsilon| }{ 4\tilde{g}^2}  \right)\right] , \label{eq:fshy}
\end{equation}  \addtocounter{subequation}{1}
where, again, $\tau_\pi = \pi \hbar/J$.
Here, the superscript `s' refers to the slow dephasing regime. 

In Fig.~\ref{fig:Vdetuning}, we plot the limiting behavior corresponding to Eq.~(\ref{eq:fshy}), together with exact numerical results.
We see that $f^\text{s}$ approximates the numerical results very well at smaller values of $|\epsilon|^{-1}$.

\subsection{Optimal Fidelity of LQR}
As discussed in the main text, the numerical results for the infidelity at fixed tunnel coupling exhibit a minimum as a function of detuning $\epsilon$, as observed in Fig.~\ref{fig:Vdetuning}.
For small $|\epsilon|^{-1}$
the curves are well-approximated by $f^\text{s}$ in Eq.~(\ref{eq:fshy}), while 
for large $|\epsilon|^{-1}$
the curves are well-approximated by $f^\text{f}$ in Eq.~(\ref{eq:fchy}).
Phenomenologically, we observe that the whole range of the numerical infidelity is well-approximated by the sum of the two limiting behaviors, so that
\begin{equation} \label{eq:fhysum}
f(\epsilon) \simeq f^\text{s}(\epsilon)+f^\text{f}(\epsilon)-1 .
\end{equation}  \addtocounter{subequation}{1}
This approximation is plotted as solid curves in Fig.~\ref{fig:Vdetuning}.

The optimal value of the detuning can be computed as the intersection point of the two infidelity curves.
Expanding Eqs.~(\ref{eq:fchy}) and (\ref{eq:fshy}) in small values of the arguments of the exponential functions, we obtain
\begin{equation} \label{eq:epoptLQR}
|\epsilon^\text{opt}| \simeq \tilde{g}\sqrt{8\Gamma/\gamma} .
\end{equation}  \addtocounter{subequation}{1}
From Eq.~(\ref{eq:fhysum}) we then obtain 
\begin{equation} \label{eq:foptLQRpre}
f^\text{opt} \simeq 1-(\pi\hbar/\tilde{g})\sqrt{\Gamma\gamma/2} .
\end{equation}  \addtocounter{subequation}{1}
For the limiting, fully mixed state of the three-level system, we should have $f=1/3$.  
Equation~(\ref{eq:foptLQRpre}) can be rewritten to reflect this limit as follows:
\begin{equation} \label{eq:foptLQR}
f^\text{opt} \simeq \frac{1}{3}+\frac{2}{3} e^{-(3h/4\tilde{g})\sqrt{\Gamma\gamma/2}} .
\end{equation}  \addtocounter{subequation}{1}
For the secondary resonance, $\tilde{g}=g_0$.
For the primary resonance, with $\tilde{g}=g_0/2$, we have
\begin{equation}
f^\text{opt} \simeq \frac{1}{3}+\frac{2}{3} e^{-(3h/2g_0)\sqrt{\Gamma\gamma/2}} .
\end{equation}  \addtocounter{subequation}{1}

As described in the main text, when certain physical constraints apply to the tunnel coupling, it may be possible to improve the gate fidelity by working \emph{away} from the optimal point, such that $\epsilon <\epsilon^\text{opt}$.
The improved fidelity is then located on one of the $f^f$ curves on the right-hand-side of Figs.~\ref{fig:Vdetuning}(a) or (b), as described by Eq.~(\ref{eq:fchy2}).

\begin{figure}
\begin{center}
\centerline{\includegraphics[scale=0.32]{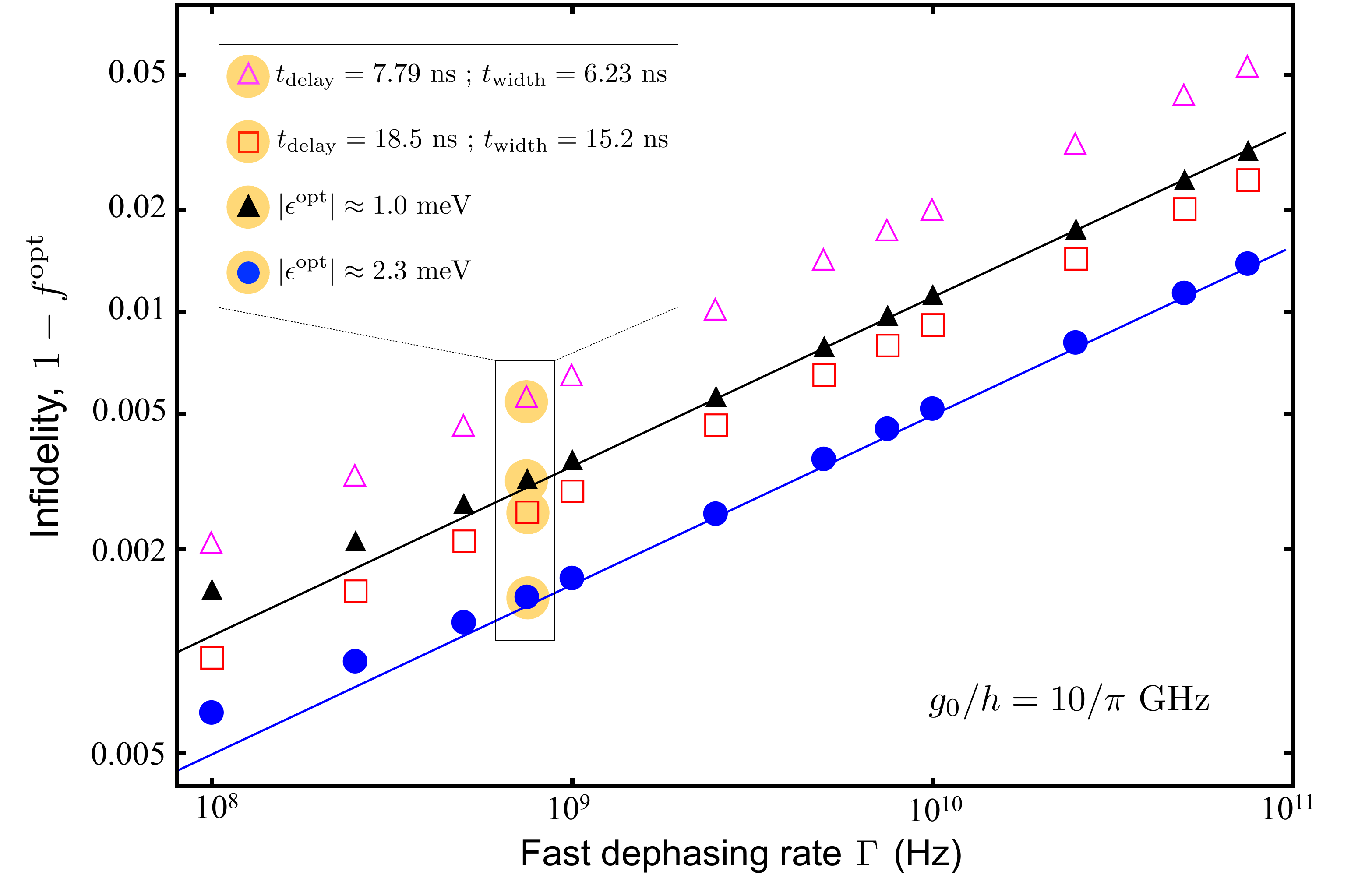}}
\caption{\label{fig:infidvsGamma}
Optimized infidelities (1-$f^\text{opt}$) vs.\ the charge dephasing rate $\Gamma$ for the LQR and STIRAP gating protocols, holding the following quantities fixed: the tunnel coupling 
$g_0/h=10/\pi$~GHz, 
and the pure dephasing rates $\gamma_\text{ST} = 2\times 10^{5}~\mathrm{Hz}$ and $\gamma_\text{hy} = 1\times 10^{6}~\mathrm{Hz}$. 
Numerical results are shown for the LQR gate, for the hybrid qubit (solid black triangles) and the ST qubit (solid blue circles). 
The solid lines show the corresponding analytical estimates of Eq.~(\ref{eq:foptLQR}) for the two cases.
Numerical results are also shown for the STIRAP protocol, for the hybrid qubit (open triangles) and the ST qubit (open squares). 
We see that the STIRAP infidelities follow those of LQR over a wide range of $\Gamma$, up to a small, overall scaling factor. 
A quantitative comparison of the optimized pulse parameters for STIRAP and LQR is provided in the boxes, for the charge dephasing rate $\Gamma = 7.5 \times 10^{8}~\mathrm{Hz}$. 
}
\end{center}
\end{figure}
\addtocounter{subfigure}{1}  

\section{Analysis of STIRAP}
\subsection{STIRAP formalism}
The STIRAP procedure involves two resonant pulses, known as the Stokes ($S$) and the pump ($P$) pulses~\cite{Bergmann:1998p1003}.
The pulsing scheme is shown in Fig.~1(e) of the main text.
The Stokes pulse is modulated at the resonant frequency $\omega_{eb}$, while the pump pulse is modulated at the resonant frequency $\omega_{ea}$.
The pulses are applied in a counterintuitive sequence, with the Stokes pulse coming before the pump pulse. 

We adopt a Gaussian shape for the pulse envelopes, with
\begin{eqnarray}
g_{S}(t) = g_0 ~\mathrm{exp}\left[ -\left( \frac{t + t_\text{delay}/2}{t_\text{width}} \right)^2\right], \label{eq:gS} \\ \addtocounter{subequation}{1} 
g_{P}(t) = g_0 ~\mathrm{exp}\left[ -\left( \frac{t - t_\text{delay}/2}{t_\text{width}} \right)^2\right]. \label{eq:gP} 
\end{eqnarray} \addtocounter{subequation}{1} 
As for LQR, we assume that the same signal is applied to the tunnel couplings between the states $|a\rangle$ and $|e\rangle$ as between the states $|b\rangle$ and $|e\rangle$; in other words, $g(t)=g_1(t)=g_2(t)$.  
Moreover, the Stokes and pump pulses are both included in $g(t)$, with
\begin{equation}\label{eq:STIRAPst}
g(t) = g_S(t) \left[1+\mathrm{cos}(\omega_{eb}t)\right]+ g_P(t)\left[1+\mathrm{cos}(\omega_{ea}t)\right] .
\end{equation}  \addtocounter{subequation}{1}

For the STIRAP protocol, we adopt a different transformation for the interaction picture with the definitions
\begin{equation}
H_0 = \begin{pmatrix}
\hbar\omega_{a} & 0 & 0 \\
0 &\hbar \omega_{b} & 0 \\
0 & 0 & \hbar \omega_{e}
\end{pmatrix} 
\quad\quad\text{and}\quad\quad
V(t) = \begin{pmatrix}
0 & 0 & g(t) \\
0 & 0 & -g(t) \\
g(t) & -g(t) & 0
\end{pmatrix}. 
\end{equation}  \addtocounter{subequation}{1}
Applying the RWA, we obtain
\begin{equation}
H^I=\begin{pmatrix} 
0 & 0 & \frac{1}{2}g_P(t) \\
 0 & 0 & -\frac{1}{2}g_S(t) \\
\frac{1}{2} g_P(t) & -\frac{1}{2}g_S(t) & 0
\end{pmatrix} . \label{eq:HISTIRAP}
\end{equation}  \addtocounter{subequation}{1}
Note that the only time dependence in $H^I$ appears in the pulse envelopes;
the resonant oscillations have been suppressed by the RWA.
The individual pulse envelopes $g_P(t)$ and $g_S(t)$ are associated with different tunneling processes, because of their distinct resonant conditions.

Similarly to LQR, the RWA is valid when many resonant oscillations occur inside a pulse envelope.
The Stokes pulse generates a $\pi$-rotation from $|b\rangle$ to $|e\rangle$ (or vice versa) when $\int_{-\infty}^{\infty}g_S(t)dt=\pi\hbar$, yielding the relation $t_\text{width}=\sqrt{\pi}\hbar/g_0$.  
(Similar considerations apply to the probe pulse.)
The characteristic width of the pulse is $\tau_\pi=\sqrt{2}t_\text{width}$, while the time needed for a full resonant oscillation is $\tau_r=2\pi/\omega_{eb}$.
The RWA requirement that $\tau_r\ll\tau_\pi$ can then be expressed as $g_0\ll2\hbar\omega_{eb}/\sqrt{\pi}$.
For the regime of interest, this is equivalent to $g_0\ll2|\epsilon|/\sqrt{\pi}$.

\begin{figure}[t]
\begin{center}
\centerline{\includegraphics[scale=0.42]{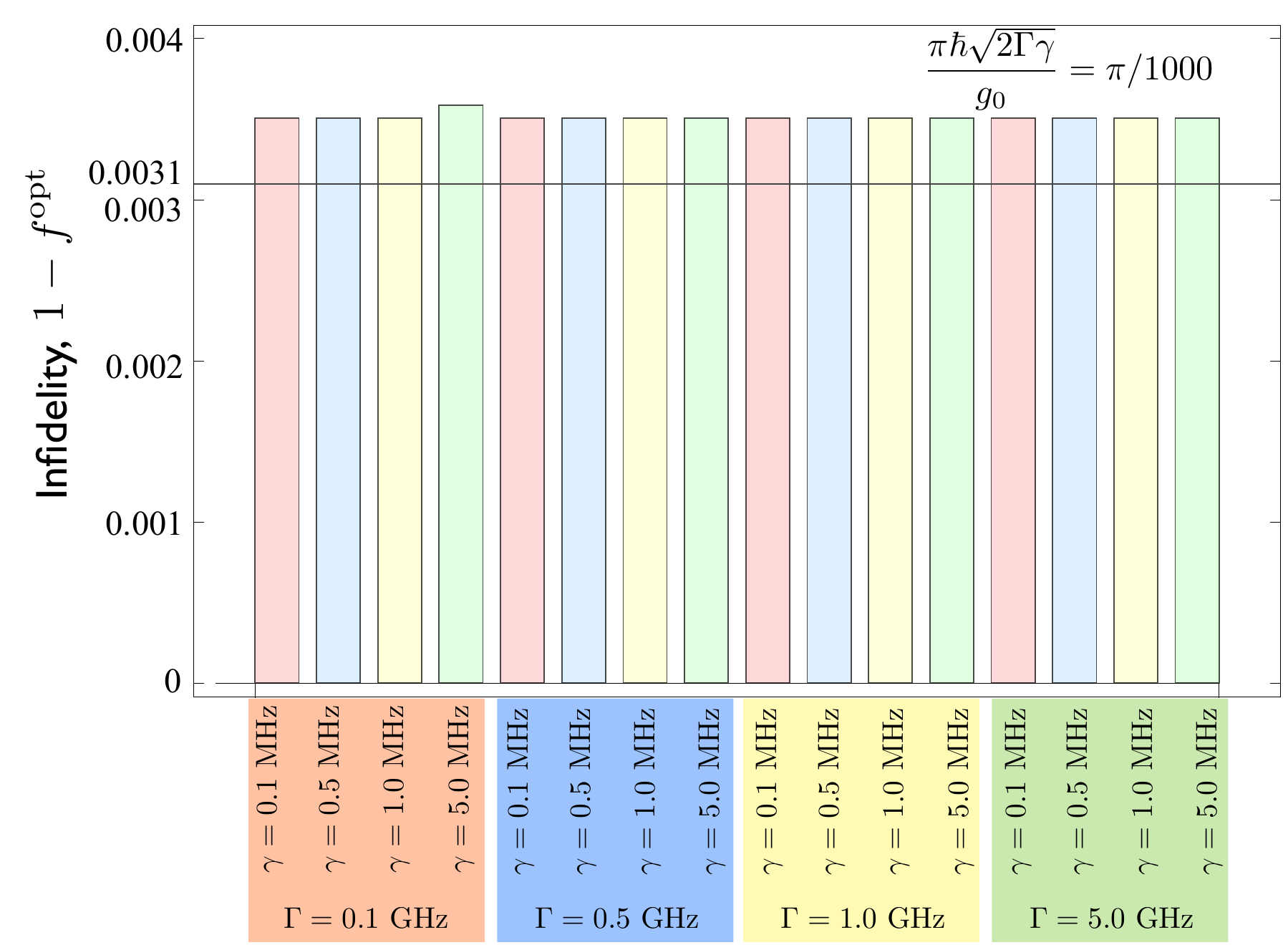}}
\caption{\label{fig:STIRAPscaling}
Scaling analysis of the optimized infidelity (1-$f^\text{opt}$), for the STIRAP protocol, for a range of $\Gamma$, $\gamma$, and $g_0$, holding the following quantity fixed:  $(\pi\hbar/g_0)\sqrt{2\Gamma\gamma_\text{hy}} = \pi/1000$ for the hybrid qubit. 
The fact that the infidelity remains constant under these conditions indicates that the scaling relation $\ln (1-f^\text{opt})\propto (\hbar/g_0)\sqrt{\Gamma\gamma}$ is satisfied for the STIRAP process.
The horizontal black lines at $0.0031$ in both panels correspond to Eq.~(\ref{eq:foptLQR}), which was derived for LQR gates.
}
\end{center}
\end{figure}
\addtocounter{subfigure}{1}  

We have noted that the STIRAP protocol described here is not a true qubit gate because it transforms $|a\rangle$ to $|b\rangle$, but not vice versa.
It is possible to achieve a true gate by implementing more sophisticated STIRAP-like pulse sequences~\cite{Lacour:2006p362}. 
For simplicity here, we consider only the sequence specified in Eqs.~(\ref{eq:gS})-(\ref{eq:STIRAPst}).
Hence, for a $\pi$-rotation, we take the initial condition to be $\rho_{aa}(0)=1$; the final process fidelity is given by $f=\rho_{bb}(\tau_\pi)$, analogous to LQR.
We solve for $\rho(t)$ using the 
master equations
given in Eqs.~(\ref{eq:masterintpic}) and (\ref{eq:DI}), with the interaction Hamiltonian given by Eq.~(\ref{eq:HISTIRAP}).
As evident from Eq.~(\ref{eq:HISTIRAP}), the STIRAP protocol does not depend directly on $\epsilon$, so 
the optimization is not over $\epsilon$ but instead is done by
obtaining an appropriate numerical relation between $t_\text{delay}$ and $t_\text{width}$.
The resulting $f^\text{opt}$ depends on $g_0$ through the relation $t_\text{width}=\sqrt{\pi}\hbar/g_0$.
Some typical fidelity results are shown in Fig.~2 of the main text.
There are no obvious approximations that could help us to find an analytical relation between $t_\text{delay}$, $t_\text{width}$, and $f^\text{opt}$.
However, there is strong numerical evidence that such a relation exists, as discussed in the next subsection.

Finally, we mention two technical details.
First, our numerical simulations were performed by truncating (chopping) the Gaussian tails in Eqs.~(\ref{eq:gS}) and (\ref{eq:gP}) 
at times $t=\pm (t_\text{delay}+t_\text{width})$, measured from the center of the double-gaussian.
Second, for the gate speed conversion shown at the top of Fig.~2 in the main text, we define the STIRAP gate speed as $(t_\text{delay}+t_\text{width})^{-1}$.

\subsection{Numerical scaling relations for the STIRAP process fidelity}
In Eq.~(\ref{eq:foptLQR}), we derived an analytical estimate for the dependence of the optimal LQR gate fidelity $f^\text{opt}$ on the tunnel coupling $g_0$, and the fast and slow dephasing rates, $\Gamma$ and $\gamma$. 
Here, we present numerical evidence that $f^\text{opt}$ for the STIRAP process follows a very similar form. 
The first piece of evidence is found in Fig.~2 of the main text, where the optimized infidelities for STIRAP and LQR are nearly coincident over a wide range of tunnel couplings $g_0$, up to a constant scaling factor of order 2. 
In that calculation, we assumed fixed dephasing rates $\Gamma$ and $\gamma$. 

Here, we perform a similar calculation of $f^\text{opt}$ as a function of $\Gamma$, keeping $\gamma$ and $g_0$ fixed, yielding the results shown in Fig.~\ref{fig:infidvsGamma}.
We again observe good agreement between STIRAP and LQR, with roughly the same scaling factor as before (\emph{i.e.}, 2). 
We also compute $f^\text{opt}$ as a function of $g_0$, $\Gamma$ and $\gamma$, in the scaling analysis shown in Fig.~\ref{fig:STIRAPscaling}.
In this case, the control parameters are all simultaneously varied while keeping the argument of the exponential function in Eq.~(\ref{eq:foptLQR}) held fixed.
Since the infidelity remains nearly constant under these conditions, we conclude that the scaling relation $\ln (1-f^\text{opt})\propto (\hbar/g_0)\sqrt{\Gamma\gamma}$ is also satisfied for the STIRAP process.

\begin{figure}
\begin{center}
\centerline{\includegraphics[scale=0.24]{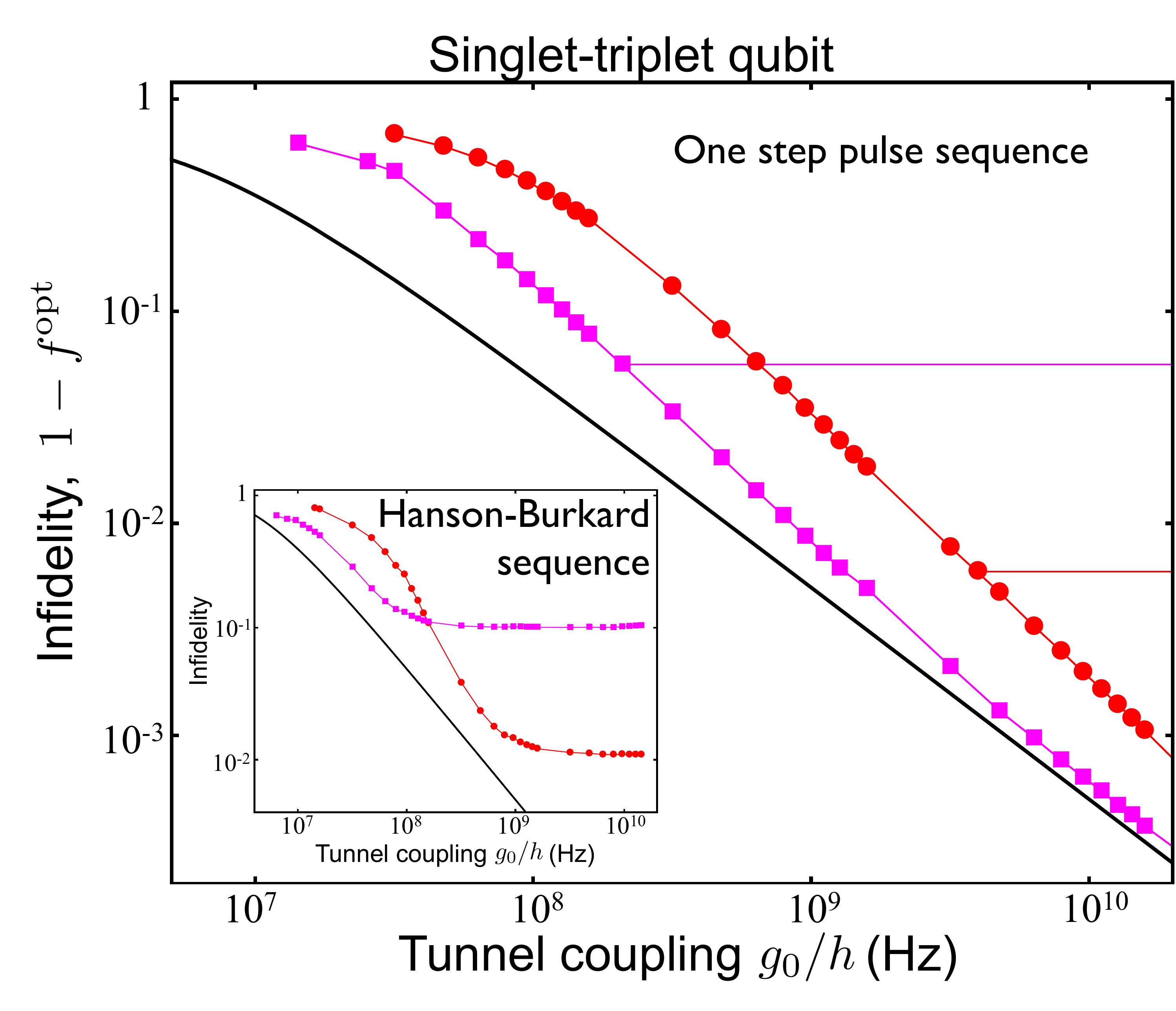}}
\caption{\label{fig:DCpulsedgating}
Investigation of the Hanson-Burkard three-pulse sequence for correcting the misorientation
of the $x$-rotation axis~\cite{Hanson:2007p050502}.
Main panel:
Optimized infidelity (1-$f^\text{opt}$) versus tunnel coupling, for a one-step pulsed gate in an ST qubit.
The results are obtained for isotopically purified $^{28}$Si, assuming two different interdot magnetic field differences:  $\Delta B=0.03~\mathrm{mT}$ (magenta squares), and $\Delta B=0.3~\mathrm{mT}$ (red circles).
The solid colored lines connect the markers.
The horizontal colored lines show the fidelities of the corresponding $z$-rotations, for the same $\Delta B$ values.
The bold black curve shows the upper bound on $f^\text{opt}$ for $x$-rotations, from Eq.~(\ref{eq:foptLQR}), which
is achieved in the limit $\Delta B\rightarrow 0$.
Inset:
Calculated infidelity for the same system and the same $\Delta B$ values, using the Hanson-Burkard (HB) three-step pulse sequence~\cite{Hanson:2007p050502} to correct for the misorientation of the $x$-rotations.
The solid black curve again shows the upper bound on $f^\text{opt}$ for $x$-rotations, from Eq.~(\ref{eq:foptLQR}).
The  HB procedure does not improve the fidelity significantly over that obtained using
the one-step procedure, and the HB fidelity
plateaus at large $g_0$ as it approaches the fidelity limit of the $z$-rotations.
}
\end{center}
\end{figure}

\section{Pulsed gating}
In typical pulsed gating implementations~\cite{Petta:2005p2180}, the tunnel couplings are held fixed while $|\epsilon(t)|$ is suddenly pulsed from the far-detuned regime (corresponding to a $z$-rotation) to a much smaller value (corresponding to an $x$-rotation).
In our simulations, the detuning pulses are assumed to occur instantaneously.  
As consistent with our AC gating analyses, we adopt $\tilde{g}=g = g_1=g_2=g_0$ for both the hybrid and ST qubits.  
 
For pulsed gates, we work in the Sch\"{o}dinger picture.
The master equation is given by
\begin{equation}\label{eq:masterSchro}
\dot{\rho} (t) = -\frac{i}{\hbar} [ H(t) , \rho(t) ] - D ,
\end{equation}  \addtocounter{subequation}{1}
where 
\begin{equation} \label{eq:HDSchro}
H = \begin{pmatrix}
-\Delta E /2 & 0 & \tilde{g} \\
 0 & \Delta E /2  & -\tilde{g} \\
 \tilde{g} & -\tilde{g} & -\epsilon(t)
\end{pmatrix} \quad\quad \text{and} \quad\quad
D =  \begin{pmatrix}
0 & \gamma \rho_{ab} & \Gamma \rho_{ae} \\
\gamma \rho_{ba} & 0 & \Gamma \rho_{be} \\
\Gamma \rho_{ea} & \Gamma \rho_{eb} & 0 
\end{pmatrix} .
\end{equation}  \addtocounter{subequation}{1}
In the limit $\Delta E\rightarrow 0$, these equations are identical in form to Eqs.~(\ref{eq:Hrwa1half}), (\ref{eq:masterintpic}), and (\ref{eq:DI}), which describe the dynamical evolution of the LQR gate.
In this limit, the $x$-rotation is therefore optimized by choosing $\epsilon^\text{opt}$ according to Eq.~(\ref{eq:epoptLQR}), with the resulting $f^\text{opt}$ given by Eq.~(\ref{eq:foptLQR}).
In other words, for $\tilde{g}=g_0$, the $\Delta E\rightarrow 0$ limit for pulsed gates gives the same optimized results as the secondary resonance for LQR gates.

The presence of a nonzero energy splitting between the qubits ($\Delta E>0$) produces a phase difference between the qubits (\emph{i.e.}, a $z$-rotation); it does not generate $x$-rotations.
The effect of including $\Delta E>0$ in Eq.~(\ref{eq:HDSchro}) is therefore to add a small $z$-component to the $x$-rotation; the latter is governed by the effective exchange coupling between the qubits, $J=2\tilde{g}^2/|\epsilon|$.
Hence, $\Delta E>0$ can only reduce $f^\text{opt}$ below the value given in Eq.~(\ref{eq:foptLQR}), due to the misorientation of the rotation axis away from $\hat{x}$.
This suppression of $f^\text{opt}$ is demonstrated in the main panel of Fig.~3 in the main text, which the same as the main panel of Fig.~\ref{fig:DCpulsedgating}.

The energy splitting $\Delta E$ is nonzero for all real devices.
Indeed, for hybrid qubits, $\Delta E$ is large enough that a one-step pulse sequence is 
untenable.
In that case, a different type of five-step pulse sequence has been proposed, as discussed in Ref.~\cite{Koh:2012p250503} and the main text.
For ST qubits, $\Delta E=g\mu_B\Delta B$ is much smaller than for hybrid qubits.
In this case, the formal definition of ``small" is $\Delta E\ll J$.
For an optimized gate with $\epsilon=\epsilon^\text{opt}(\tilde{g})$, this can be rewritten as $\Delta E\ll \tilde{g}\sqrt{\gamma/2\Gamma}$, which can be quite restrictive, particularly for GaAs devices.

The problem of misoriented $x$-rotations is well known, and alternative gating schemes have been proposed as a solution~\cite{Levy:2002p147902}.
Hanson and Burkard (HB) have proposed a specific three-step pulse sequence that uses a combination of $x$ and $z$-rotations~\cite{Hanson:2007p050502}.
In their scheme, the $x$-like component is split in half, and an intermediate $z$-rotation is inserted to compensate for the $x$ misorientation.

A potential problem with the HB proposal is that the three-step procedure could inherit the underlying flaws of both the $x$ and $z$ components of the protocol.
Indeed, this is what we find in our simulations.
We have implemented the HB sequence for pulsed gates in ST qubits, as shown in the inset of Fig.~\ref{fig:DCpulsedgating}.
For small $g_0$ and large $\Delta E$, the HB procedure provides
 slight improvements in fidelity.
 However,  $f^\text{opt}$ never reaches the upper bound suggested by Eq.~(\ref{eq:foptLQR}),  
and for large $g_0$, the optimized fidelity is limited by the fidelity of the $z$-rotation component.  
(This is evident by comparing the inset with the main panel in Fig.~\ref{fig:DCpulsedgating}.)
Overall, the HB procedure is not found to give significant improvements in the gate fidelity.

{\color{black}
Finally, we comment on other possible sources of error associated with pulsed-$\epsilon$ gates.
(1) In the simulations performed here, we assumed perfect (instantaneous) square wave pulses.
However, the pulses used in real quantum dot experiments have finite rise times, due to filtering and other hardware limitations.
As discussed in the main text, the criterion for satisfying the sudden approximation is given by $\hbar \, (d\Delta/dt) \gg g^2$~\cite{Vutha:2010p389}. 
We note that finite rise times are more limiting for hybrid qubits, because of their relatively large tunnel couplings.
Such limitations are purely hardware related.
(2) Pulsed gates in hybrid qubits involve pulses to two different energy level anticrossings.
The operations at each anticrossing are accurate when they are distinct and separate.  
When the qubit energy level splitting $\Delta E_{01}$ is small, or the tunnel coupling $g$ is large, this condition will not be satisfied.
In this case, the system evolution is coherent, but potentially complicated, and difficult to control.
The inset of Fig.~3 in the main text shows reduced fidelity due to this effect.
Another complication of pulse gating of hybrid qubits arises because of
additional energy level anticrossings that
may be present~\cite{Shi:2013preprint}.
If these anticrossings are not distinct and separate, then pulsed gating can cause other non-qubit states to be populated.
Such effects can be addressed by appropriate pulse-shaping.
For simplicity, we have not considered such leakage effects in the calculations reported here.
(3) Low-frequency noise due to the motion of surface charge and other defects in the semiconductor causes uncertainty in the values of $\epsilon$ and $g$ during pulsed gate operations, and ultimately suppresses the  fidelity.
In this work, we have focused on the uncertainty in the tunnel coupling $g$, which causes dephasing when we implement exchange interactions~\cite{Barrett:2002p125318}.
Uncertainty in the detuning $\epsilon$ also causes errors.
The errors are most prominent near $\epsilon=0$ because $\partial J/\partial \epsilon$ is large, and
they are relatively unimportant in the far-detuned regime ($\epsilon \ll 0$) where $\partial J/\partial \epsilon$ is small~\cite{Taylor:2007p035315}.
Our simulations indicate that the highest gate fidelities are obtained in the far-detuned regime, where $\epsilon$ errors are suppressed.}

\end{onecolumngrid}



\begin{thebibliography}{10}
\bibitem{footnote2}
The standard, Stokes-pump STIRAP protocol yeilds $|a\rangle\rightarrow |b\rangle$ but not $|b\rangle\rightarrow |a\rangle$, and is therefore not an $x$-rotation.
\end{thebibliography}
\end{document}